\normalfont
\documentclass[prb,twocolumn,showpacs,preprintnumbers,amsmath,amssymb]{revtex4}

\usepackage{graphicx}
\usepackage{dcolumn}
\usepackage{bm}
\usepackage[usenames]{color}
\usepackage{multirow}
\usepackage[normalem]{ulem}

\usepackage{color}

\allowdisplaybreaks[1]

\graphicspath{{Figs/}}

\begin{document}


\title{Impact of local arrangement of Fe and Ni on the phase stability and magnetocrystalline anisotropy in Fe-Ni-Al Heusler alloys}

\author{Vladimir~V.\ Sokolovskiy$^{1}$}
\author{Olga~N.~Miroshkina$^{2,1}$}
\author{Vasiliy~D.\ Buchelnikov$^{1}$}
\author{Markus~E.~Gruner$^{2}$}


%
\affiliation{$^{1}$Faculty of Physics, Chelyabinsk State University, Chelyabinsk 454001, Russia}
\affiliation{$^{2}$Faculty of Physics and Center for Nanointegration, CENIDE, University of Duisburg-Essen, 47048 Duisburg, Germany}

\date{\today}

\begin{abstract}
On the basis of density functional calculations, we report on a comprehensive study of the influences of atomic arrangement and Ni substitution for Al on the ground state structural and magnetic properties for Fe$_2$Ni$_{1+x}$Al$_{1-x}$ Heusler alloys. We discuss systematically the competition between five Heusler-type structures formed by shuffles of Fe and Ni atoms and their thermodynamic stability. 
 All~Ni-rich Fe$_2$Ni$_{1+x}$Al$_{1-x}$ tend to decompose into a dual-phase mixture consisting of Fe$_2$NiAl and~FeNi. The~successive replacement of Ni by Al leads to a change of ground state structure and eventually an increase in magnetocrystalline anisotropy energy~(MAE).
 We predict for stoichiometric Fe$_2$NiAl a ground state structure with nearly cubic lattice parameters but alternating layers of Fe and Ni possessing an uniaxial MAE which is even larger than tetragonal L1$_0$-FeNi. 
 This opens an alternative route for improving the phase stability and magnetic properties in FeNi-based permanent magnets.

%
%
\end{abstract}


\maketitle

\section{Introduction}

The~demand for permanent magnets for modern application technology is very high.
Permanent magnets are used in automating and robotics industry and constitute a parts of an electric vehicles, wind turbines, drives, and storage and magnetic cooling devices~\cite{Skomski-2016,Skomski-2016b,Hono-2018,Skokov-2018,Mohapatra-2018,Kovacs-2020,Coey-2020}. 
Application grade materials should exhibit particular intrinsic properties, in first line
 a high saturation magnetization~($M_s$) in combination with a large uniaxial anisotropy and a sufficiently high Curie temperature~($T_C$). 
But also extrinsic properties must match, such as a micro/nanostructure that preserves coercivity, high remanent magnetization and a high magnetic flux density in combination with mechanical stability and corrosion resistance. Thermodynamic stability and ductility which affect the processing of the material are also important demands.

Nowadays, the most widespread high-performance permanent magnets are based on rare earth elements.
However, such materials are extremely expensive and finding inexpensive analogs consisting of abundant elements with similar performance is one of the most urgent and challenging tasks in modern materials science~\cite{Kuzmin-2014,Mccallum-2014,Niarchos-2015,Hirosawa-2015,Skomski-2016,Li-2016,Skokov-2018}. 
One~of the most intensely discussed candidate is equiatomic FeNi~\cite{Edstrom-2014,Werwinski-2017,Cui-2018, Tian-2019, Tian-2020, Tian-2021, Tuvshin-2021} in the tetragonally ordered L1$_0$ structure (tetrataenite). 
It~is characterized by MAE of 
$\approx 0.32 - 1.3$~MJ/m$^3$ (Refs.~\onlinecite{Pauleve-1962,Pauleve-1968, Poirier-2015}) and might be considered as a rare-earth free competitor of Nd$_2$Fe$_{14}$B  ($3.7-5.15$~MJ/m$^3$, see Refs.~\onlinecite{Skomski-2019permanent,Pastushenkov-2005,Otani-1987})
, but hampered by a much smaller reduction with increasing temperature~\cite{Lewis-2014inspired}.

However, up to date, the synthesis of L1$_0$ ordered FeNi is very challenging, 
because it becomes thermodynamically stable at temperatures lower than the order-disorder transition temperature, which is in the range of~$200-300$~$^{\mathrm{o}}$C. 
This~leads to an extremely slow atomic diffusion.
Indeed, natural tetrataenite is found in meteorites with a slightly Fe-rich (Fe:~$50.47\pm1.98$~at.\% and Ni: $49.6\pm 1.49$~at.\%) composition and a body centered tetragonal ({\it bct}) structure with alternating layers of Fe and Ni along the elongated $c$-axis~\cite{Cui-2018, Lewis-2014inspired}.
It~has been shown, that it is possible to grow L1$_0$-FeNi in the laboratory as thin films irradiated with neutrons below 593~K~\cite{Neel-1964}, in a low temperature cyclic
oxi-reduction process~\cite{Lima-2001}, an alternate monatomic layer deposition technique~\cite{Shima-2007},  crystallizing an amorphous alloy with equiatomic composition at a crystallization temperature close to the order-disorder transition temperature~\cite{Makino-2015}.
The preparation of L1$_0$-FeNi films on Au-Cu-Ni buffer-layer was studied by Giannopoulos~\textit{et~al.}~\cite{Giannopoulos-2018}.
With their high-throughput magnetic characterization methods the authors revealed the presence of a hard magnetic phase with an out-of-plane anisotropy, while the MAE density increases from 0.12 to 0.35~MJ/m$^3$.
Goto~\textit{et~al.}~\cite{Goto-2017} suggested  to synthesize single-phase L1$_0$-FeNi powder with a high degree of order through nitrogen insertion and topotactic extraction.

Despite these efforts, bulk tetrataenite has not been obtained on a large scale yet.
Therefore, scientist keep on trying to find new ways to overcome the obstacles in producing L1$_0$-FeNi.
Tian~\textit{et~al.} worked on the optimization of the tetrataenite phase in terms of
the order-disorder transformation~\cite{Tian-2019}, the effect of pressure~\cite{Tian-2020} and alloying~\cite{Tian-2021} by non-magnetic (Al and Ti) or magnetic (Cr and Co) atoms.
The effect of alloying was also investigated by other authors: Tuvshin~\textit{et~al.}~\cite{Tuvshin-2021} showed that the addition of B can improve both the phase stability and increase MAE up to 2.6~MJ/m$^3$.
Izardar~\textit{et~al.}~\cite{Izardar-2020} studied the interplay between chemical order and the magnetic properties and found that the reduction of chemical long-range order by 25\% does not reduce the~MAE significantly.
However, the central issues of proper growth conditions that stabilize the L1$_0$ phase are still not fully resolved.

Over the past decades, also full Heusler alloys of $X_2YZ$ stoichiometry were considered for permanent magnet applications. 
The~properties of these alloys can be tuned flexibly depending on composition and structural defects. 
Recent theoretical studies~\cite{Matsushita-2017,Herper-2018,Gao-2020} 
suggested Heusler alloys based on Ni~\cite{Herper-2018, Gao-2020}, Fe~\cite{Gao-2020, Matsushita-2017}, Co~\cite{Gao-2020, Matsushita-2017}, Rh, Au, and Mn~\cite{Gao-2020} as promising classes of hard magnetic materials.

Recently Herper~\cite{Herper-2018} revealed the effect of $Y$, $Z$, and fourth element doping as well as lattice deformation on MAE for Ni$_2YZ$ ($Y=$~Mn, Fe, Co and $Z=$~B, Al, Ga, In, Si, Ge, Sn).
It~was found that the phase stability and MAE of Ni$_2$Co$Z$ increases by adding a fourth element from the main groups~III and IV to the $Z$ sublattice, while the addition of Fe to the $Y$ sublattice does not have a similar effect. 
Besides, small deviations from cubic structure lead to a doubling of the MAE in Ni$_2$FeGe. 
For~most systems 
considered in Ref.~\onlinecite{Herper-2018}
, the MAE exhibits a quasilinear dependence on the tetragonality of the lattice and changes its sign around~$c/a=1$. 
However, for Ni$_2$FeGe, the MAE remains uniaxial for the $c/a$ from 0.85 to~1.45. 
In~this case, a small deviation from $c/a=1$ leads to an increase in MAE from 1 to 2~MJ/m$^3$. 

Gao~\textit{et~al.}~\cite{Gao-2020} reported on the influence of interstitial atoms (H, B, C, and N) on the MAE for~Fe-,~Ni-,~Co-,~Rh-, ~Au-, and Mn-based Heusler alloys. 
Among tetragonal structures, authors found 32 compositions with high energy of uniaxial anisotropy ($> 0.4$~MJ/m$^3$) and 10 compositions with high energy of in-plane anisotropy. 
It~can be noted that the addition of H atoms in the interstices leads to large MAE values ($1.5$~MJ/m$^3$). 
For~Ni$_2$FeGa, MAE also increases from 0.23~MJ/m$^3$ to 1.43, 0.94, and 0.56~MJ/m$^3$ with the addition of interstitial  N, C, and H atoms, respectively. 
In~contrast, MAE decreases with the addition of interstitial atoms in Fe$_2$CoGa.

Fe$_2 YZ$ ($Y =$~Ni, Co, Pt) and Co$_2 YZ$ ($Y =$~Ni, Fe, Pt), where $Z =$Al, Ga, Ge, In, Sn for both series,  were considered by Matsushita~\textit{et~al.}~\cite{Matsushita-2017}. 
Among the 30 investigated compositions, the cubic phase is found to be favorable in 15~only. 
In particular, for Co$_2$Ni$Z$, Co$_2$Pt$Z$, and Fe$_2$Pt$Z$, as well as for Fe$_2$NiGe and Fe$_2$NiSn, tetragonal structures prevail. 
For~15 tetragonal structures, the MAE is in the range from $-12$~MJ/m$^3$ for Co$_2$PtAl to $5.19$~MJ/m$^3$ for Fe$_2$PtGe.
In~case of compounds without Pt, the values of the MAE are significantly lower: from $-2.38$~MJ/m$^3$ for Co$_2$NiGa to 1.09~MJ/m$^3$ for Fe$_2$NiSn. 
Note, for magnetically hard materials, negative values correspond to moments oriented perpendicular to the tetragonal axis, leading to an easy plane, whereas positive MAE values denote an easy axis which is the case of interest for application. 
In~this regard, the authors have shown that among the considered alloys, MAE is positive in the case $Z$ element is Ge or~Sn.
It is important to note that the above mentioned theoretical studies considered the regular L2$_1$ and inverse Heusler structures.

As FeNi represents one corner of the ternary Fe-Ni-Al phase diagram, these systems may be of interest as potential permanent magnets, as well.
However, so far mainly stoichiometric Fe$_2$NiAl is reported in the literature, 
which does not provide a uniform picture.
Theoretical studies~\cite{Zhang-2013,Gupta-2014,Dahmane-2016,Matsushita-2017} suggested that Fe$_2$NiAl possesses the inverse cubic structure (Hg$_2$CuTi prototype), which confirms part of the experimental results~\cite{Popiel-2004,Zhang-2013}, but stands in conflict with others. 
According to the direct synthesis calorimetry of standard formation enthalpies conducted by Yin~\textit{et~al.}~\cite{Yin-2015}, Fe$_2$NiAl crystallizes in a B2 structure in agreement with the high temperature phase diagram~\cite{Zhang-2009}. 
Other reports suggest, that Fe$_2$NiAl alloys are characterized by a solid solution decomposition into two isomorphous body centered cubic (\textit{bcc}) phases, consisting of Fe-rich particles ($\beta$-phase) and a weak-magnetic NiAl-based matrix ($\beta_2$-phase)~\cite{Bradley-1937,Menushenkov-2017} and
as~a result of the decomposition, a miscibility gap appears on the phase diagram at lower temperatures~\cite{Hao-1984,Zhang-2008,Zhang-2009}.
However, subsequent heat treatment can improve the hard magnetic properties of Fe-Ni-Al systems~\cite{Buschow-2007,Menushenkov-2015a,Menushenkov-2015b}. 
In~the view, that the question about the type of ordering in Fe$_2$NiAl is still unsettled,
it~is of fundamental interest to consider further types of ordering motifs and extend this investigation to the Ni-excess compositions with FeNi as the end point.
In~this paper, we therefore present a systematic first-principle study of structural, magnetic, and electronic properties of Fe$_2$Ni$_{1+x}$Al$_{1-x}$ ($x =$~0, 0.25, 0.5, 0.75, and 1) in dependence on a competing crystal ordering in austenite and martensite phases. 
The~paper is organized as follows: After this introduction, 
Section~II is devoted to the technical details of the \textit{ab~initio} calculations. 
In~Section~III, the results of our calculations are presented and discussed, with emphasis on the energetic order of along the tetragonal transformation path in Sec.~IIIA, the stability of the cubic structure of Fe$_2$NiAl against decomposition in Sec.~IIIB and a comparison of the magnetic properties including MAE of the most stable structures in  Sec.~IIIC. 
Section~IIID reports exchange constants and Curie temperatures, while the stability of off-stoichiometic at finite temperatures is finally discussed in Sec.~IIIE.   
Concluding remarks are given in Sec.~IV.

\section{Computational Details}
Structural, magnetic, and thermodynamic properties of Fe-Ni-Al were calculated from first-principles using the plane-wave basis set and the projector augmented wave~(PAW)  method implemented in the Vienna \textit{ab~initio} simulation package\cite{Kresse-1996} (VASP).
For the exchange and correlation functionals, the generalized gradient approximation~(GGA) in the scheme of Perdew, Burke, and Ernzerhof ~(PBE)~\cite{Perdew1996} was applied; 
we employed~PAW potentials with the following valence states: 3$p^6$3$d^7$4$s^1$ for Fe, 3$d^8$4$s^1$ for Co, 3$p^6$3$d^8$4$s^2$ for Ni, and 3$s^2$3$p^1$ for~Al.
The~kinetic energy cut-off for the augmentation charges was set to 1000~eV while the energy cut-off for plane wave basis set was taken of~750~eV. 
The~Brillouin zone integration was performed using the first order method of Methfessel-Paxton 
with a smearing parameter of  0.1\,eV
on a uniform Monkhorst-Pack $15\times15\times15$ $k$-point grid. 
Electronic selfconsistency was assumed when the total energy difference between two consecutive steps reached $10^{-7}$~eV. 

\begin{figure}[t!!] 
%
\centerline{
\includegraphics[width=\columnwidth,clip]{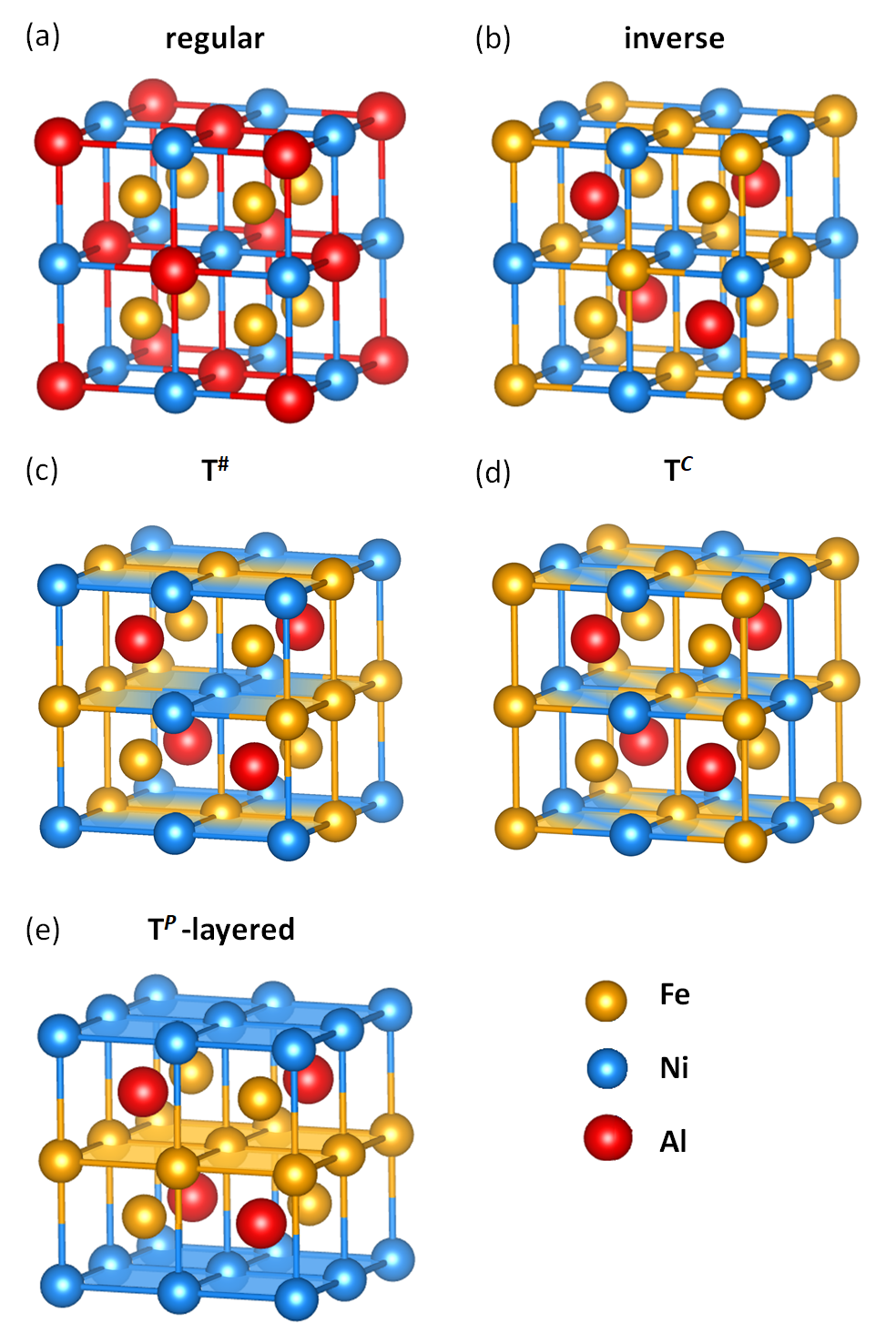}}
\caption{
The five considered  structures for the Fe$_2$NiAl Heusler alloy with cubic lattice parameters: 
(a)~Cu$_2$MnAl-type Regular Heusler structure, 
(b)~Hg$_2$TiCu-type inverse Heusler structure, 
(c)~T$^\#$, (d)~T$^c$, and (e)~T$^p$ structures. 
T$^\#$, T$^p$, and T$^c$-structures can be obtained from the inverse structure by various atomic shuffles. 
For the T$^p$ and T$^c$ structures, originally suggested by Neibecker~\textit{et al.}\cite{Neibecker-2017}, the Ni atoms and part of Fe atoms are ordered into alternating planes and columns, correspondingly. 
T$^\#$~can be conceived as chains of Fe and Ni atoms, which are arranged perpendicular to each other in alternating planes.
}
\label{Fig-1}
%
\end{figure}

The alloys were modeled in a 16-atom cubic supercell, 
which represents off-stoichiometric compositions Fe$_2$Ni$_{1+x}$Al$_{1-x}$ with concentration steps of~$x=0.25$, while still allowing for extensive calculations of phonon dispersions and the accurate determination of the MAE~
(see the supplemental Material, SM, in Ref.~\onlinecite{SM}).
The~atomic positions for the distorted configurations were considered fully relaxed when interatomic forces were smaller than~10$^{-2}$~eV/\AA.

We considered five types of ordered atomic arrangement depicted in Fig.~\ref{Fig-1}.
In~the regular L2$_1$ structure (space group $Fm\overline{3}m$, No.~225, prototype Cu$_2$MnAl),  Fe atoms locate at equivalent 8$c$ (1/4, 1/4, 1/4) and (3/4, 3/4, 3/4) sites while Al and Ni atoms occupy 4$a$ (0, 0, 0) and 4$b$ (1/2, 1/2, 1/2) sites, see Fig.~\ref{Fig-1}(a). 
For~the inverse (FeNi)FeAl Heusler structure (space group $F\overline{4}3m$, No.~216, prototype Hg$_2$TiCu), Fe atoms are placed at non-equivalent 4$a$  (0, 0, 0) and 4$c$ (1/4, 1/4, 1/4) sites while Ni and Al are placed at 4$b$ (1/2, 1/2, 1/2) and 4$d$ (3/4, 3/4, 3/4) positions, respectively, c.\,f.\ Fig.~\ref{Fig-1}(b).
In~addition, we considered three structures related to the inverse Heusler type, with
layer-wise and columnar ordering of Fe and Ni atoms on the 4$a$ and 4$b$ sites.
The~T$^\#$ structure is characterized by columns of Fe and Ni atoms located at 4$a$ and 4$b$ sites, which change their orientation from layer to layer as shown in Fig.~\ref{Fig-1}(c).
T$^c$~consists of alternating Fe and Ni atomic columns along the [001] direction ($z$-axis) and can also be conceived as Fe and Ni layers alternating along [110], c.\,f. Fig.~\ref{Fig-1}(d).
T$^p$ is composed of layers of Fe and Ni atoms alternating along [001], see Fig.~\ref{Fig-1}(e). 
The latter two structures were suggested in an earlier work on quaternary stoichiometric NiCoMn(Ga,Al) alloys~\cite{Neibecker-2017}, where they proved competitive to the conventional and the inverse Heusler structure in Fig.~\ref{Fig-1}(a,b), in agreement with the experimental findings.

To determine the MAE for the fully relaxed structures, we made use of the magnetic force theorem\cite{Liechtenstein-1984, Liechtenstein-1985, Liechtenstein-1987}: Starting from the wave functions obtained from accurate self-consistent collinear ground-state calculations with a scalar-relativistic Hamiltonian, we performed non-collinear calculations including the spin-orbit term coupling in a non-self-consistent scheme, where the quantization axis was rotated into two different orientations, [100] and [001], but the wave functions were kept fixed otherwise. 
The~MAE is then given by the energy difference between two spin moment directions \cite{Enkovaara-2002,Umetsu-2006,Gruner-2008,Edstrom-2015},
as
$$\Delta E_{\mathrm{MAE}} = E_{100}^{\mathrm{tot}} -  E_{001}^{\mathrm{tot}},$$ 
where $E_{001}^{\mathrm{tot}}$ and $E_{100}^{\mathrm{tot}}$ are the total energies with the respective spin orientations.
For~a positive value of MAE, the out-of-plane spin configuration (easy axis) is energetically favorable and \textit{vice versa} the in-plane spin direction (easy plane) is preferred. 
For~a deeper analysis of the dependence of the MAE on the composition in the range~$x$ from 0 to~0.25, we carried out calculations using a $2\times2\times1$ supercell containing 64 atoms. Here, a $6\times6\times6$ $k$-point mesh was taken into account. By~this means we considered Fe$_2$Ni$_{1+x}$Al$_{1-x}$ with $x = 0$, 0.0625, 0.125, 0.1875, and 0.25 in austenitic state.

Finite-temperature properties were obtained from the calculations of the vibrational density of states~(VDOS) in the harmonic approximation within the direct (force-constant) approach using the Phonopy~\cite{Phonopy} code.
Depending on the composition and the respective primitive cells, we employed $3 \times 3 \times 2$, $3 \times 3 \times 3$, $4 \times 4 \times 2$, $4 \times 4 \times 4$, and $6 \times 6 \times 4$ supercells to calculate
Hellmann–Feynman forces from a set of non-equivalent atomic displacements using the VASP code, to construct the force constant matrix of the system.
From the eigenvalues of the resulting dynamical matrix we obtained the VDOS and, finally, the lattice free energies.

Curie temperatures were determined from the mean-field approximation to Heisenberg model parameterized from {\em ab initio} data.
The~Heisenberg exchange coupling constants were obtained with the SPR-KKR package~\cite{Ebert-2011} employing the PBE exchange-correlation functional.  
The~calculations were performed for the most favorable structures in austenite and martensite based on the 16-atom supercells used in  the VASP calculations.
The~angular momentum expansion for the major component of the wave function, $l_{max}$, was restricted to $d$-states; the~energy convergence was set to 0.01~mRy.
For~self-consistent cycles, the scattering path operator was calculated by the Brilloiun-zone integration assuming the  $57^3$~\textit{k}-mesh grid with 4495~\textit{k}-points.

\section{Results and discussion}
\subsection{Cubic to tetragonal distortion}

\begin{table*}[t]
    \caption{Favorable crystal structure, lattice parameter $a$ 
    (in~\AA), tetragonal distortion ratio~$c/a$,  and formation energy~$E_{\mathrm{form}}$ (in~eV/f.u.) of Fe$_2$Ni$_{1+x}$Al$_{1-x}$ as well as the total energy difference $\Delta E_{c/a}$ (in~meV/atom) between the austenitic ($c/a\approx 1$) and martensitic states ($c/a> 1$) with the lowest energy. 
    The data are collected from the full geometric crystalline structure optimization. 
    For~the martensitic phase of Fe$_2$Ni$_{1.5}$Al$_{0.5}$, the data represent the average over individual tetragonal distortions of a 16-atom supercell along the three Cartesian axes~\cite{x05-cell}.
   }
    \centering
    \begin{ruledtabular}
        \begin{tabular}{lccccccccc}
            \multirow{2}{*}{Composition} & \multicolumn{4}{c}{Austenite} & \multicolumn{4}{c}{Martensite}  & \multirow{2}{*}{$\Delta E_{c/a}$} \\ \cline{2-5} \cline{6-9}          
             & Structure & $a_0$ & $c/a$ &   $E_{\mathrm{form}}$     & Structure & $a_t$ & $c/a$ & $E_{\mathrm{form}}$ &    \\ \hline
            $x = 0.0$  &   T$^p$   &   5.744  &  0.992    &    -1.191     & ---   &   &   &   &   \\
             
            $x = 0.25$ &   T$^p$   &   5.728 &   0.987  &   -0.799 &  ---   &   &   &   &   \\
             
            $x = 0.50$ &   Inverse &  5.699   & 0.999 &  -0.404 &   L$1_0$ &   5.004  &   1.475   &   -0.519  &   26.170  \\
             
            $x = 0.75$ &   Inverse &   5.677  &   1.000  &     -0.109 &  L$1_0$    &   5.016  &   1.446   &   -0.328   &   55.702 \\
             
            $x = 1.0$  & ---   &    &     &  \textbf{  }    &   $bct$-L$1_0$ &   5.019   &   1.424  &   -0.151   &    \\
             & ---   &    &     &  \textbf{  }    & $fct$-L$1_0$\footnote{ 
             For better comparison with the ternary compositions, we list the lattice parameters of L$1_0$-FeNi in terms of the bct tetragonal Heusler cell, and additionally with respect to the conventional (4~atoms) fct L$1_0$ cell. 
             The~lattice parameters transform as: $a_{fct}=\frac{a_{bct}}{\sqrt 2}$ and $c_{fct}=\frac{c_{bct}}{ 2}$.} 
             &   3.549   &   1.007  &    &    \\
        \end{tabular}
    \end{ruledtabular}
    \label{table-1}
\end{table*}

\begin{figure}[b!!!] 
    \centering
    \includegraphics[width=\columnwidth]{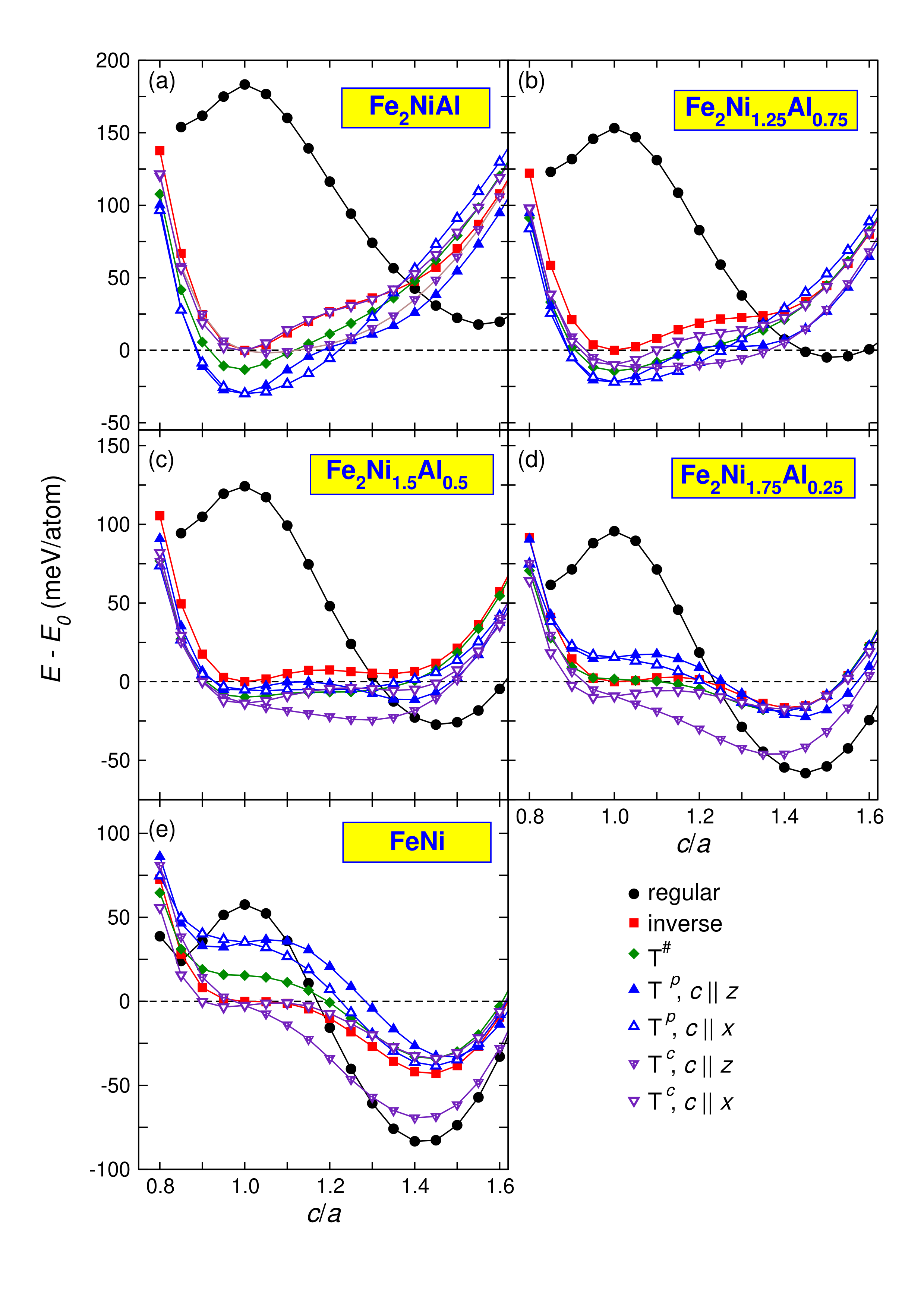} 
    \caption{
        Variation of the total energy as a function of tetragonal ratio~$c/a$ for Fe$_2$Ni$_{1+x}$Al$_{1-x}$ with various atomic order in a cubic structure: (a)~$x = 0.0$, (b)~$x = 0.25$, (c)~$x = 0.5$,  (d)~$x = 0.75$, and (e)~$x = 1.0$. 
        The~binding surface $E(c/a)$ is plotted with respect to the energy~$E_0$ of the inverse cubic structure.         }
    \label{Fig-2}
    \end{figure}

To~determine the equilibrium ground-state properties of Fe-Ni-Al, we apply a volume conserving tetragonal distortion to the fully optimized cubic 16 atom unit cell of the Heusler compound. From the calculations of the total energy as a function of the volume and the $c/a$ ratio we determined the optimized lattice constants for all compositions with the investigated crystalline structures in austenite ($c/a\approx 1$) and martensite ($c/a \neq 1$), the data of the most important ones are summarized in Table~\ref{table-1}.

Figure~\ref{Fig-2} displays the corresponding difference in total energy $E(c/a)$ as a function of the tetragonal distortion $c/a$ with respect to the cubic inverse Heusler structure for Fe$_2$Ni$_{1+x}$Al$_{1-x}$ ($x =$~0, 0.25, 0.5,  0.75, and 1.0) with the different types of atomic order shown in Fig.~\ref{Fig-1}. Note that for all cases the ferromagnetic~(FM) configuration is~found more favorable than potential ferri- or antiferromagnetic arrangements of the magnetic moments. 
As for some of the structures the cubic symmetry is already broken for $c/a=1$ -- in particular for T$^p$ and T$^c$, which are characterized by a layered atomic arrangement -- the energy landscape varies if we align the crystallographic $c$ axis along [001] (i.\,e., the Cartesian $z$-axes) or [100] ($x$-axis), while the deformation along [010] ($y$-axis) remains equivalent to [100]. The corresponding binding curves are labeled as $c||x$ and $c||z$ in Fig.~\ref{Fig-2}, which may indeed show significant differences. For realistic samples, where we may expect anti-phase boundaries between grains with different orientations of the layering (or a polycrystalline sample), energies at $c/a\neq 1$ can be averaged over for the alignment of $c$ along all three Cartesian axes.

We first focus on the stoichiometric case Fe$_2$NiAl. 
It~can be seen from Fig.~\ref{Fig-2}(a), that the layered T$^p$ structure with $c/a\approx 1$, i.\,e., without significant additional tetragonal distortion of the unit cells, turns out to be the ground state. This means that the austenite phase, characterized by $c/a=1$, remains stable down to low-temperatures and a transformation to a martensite structures with $c/a\neq 1$ is suppressed.

T$^p$ is $\approx$25~meV/atom lower in energy than T$^\#$, the inverse Heusler, and the T$^c$ structures, which are very close in the energy to the regular L2$_1$ Heusler structure. 
The~ close competition between of T$^p$, T$^\#$, T$^c$, and inverse structures hampers the chemical ordering of the samples under typical experimental conditions and makes the B2-type structure a likely outcome, rather than the inverse Heusler structure, as reported previously\cite{Zhang-2013, Popiel-2004}. However, under favorable experimental preparation conditions (applied stress, the rate of the cooling process or annealing time of a sample, etc.) it might be possible to synthesize the T$^p$ phase experimentally, for instance in epitaxial thin films by a sputtering procedure. 

A small deviation from the stoichiometry up to $x~=~0.25$ keeps the T$^p$ structure in austenite as the most favorable one, while the energy difference between other structures is reduced, which increases the competition between them. 
Apart from the regular Heusler structure, all ordering motifs of  Fe$_2$NiAl and  Fe$_2$Ni$_{1.25}$Al$_{0.75}$ remain stable against a tetragonal distortion of the unit cell, see Figs.~\ref{Fig-2}(a,b). 

In turn, for  compositions with $x \geq 0.5$, the global energy minimum is obtained around $c/a =1.45$ for the regular L1$_0$ Heusler phase, see Figs.~\ref{Fig-2}(c-e), while the binding curves $E(c/a)$ for the other structures become remarkably flat.
The global minimum of the L1$_0$ phase becomes deeper as compared to the other structures, when we approach the binary FeNi compound. 
Figs.~\ref{Fig-2}(c-e) show, that only the tetragonal T$^c$ structure is in close proximity to the regular L1$_0$ under the assumption that the tetragonal distortion is performed along $z$ axis.  In~this case, the energy difference between the two phases is about 6, 13, and 15~meV/atom for compositions with $x = 0.5$, 0.75, and 1.0, correspondingly. 
The difference between energy solutions for the T$^c$ structure with the tetragonal $c$ axis oriented along $x$ and $z$ increases significantly towards binary FeNi with $c/a \approx \sqrt{2}$.

In the binary limit, the T$^c$ structure for FeNi consists of a stacking of mixed Fe/Ni layer in a staggered arrangement in contrast to the ordered L1$_0$ phase, in which Fe and Ni alternate layer-wise, and these ordering motifs cannot be transformed into each other by a simple site swapping process.
Therefore, the close competition between T$^c$-type FeNi distorted along~$z$ and L1$_0$ tetrataenite might be a reason for the difficulties to synthesize L1$_0$-FeNi with a high degree of order in experiment. 

For compositions between $x=0.5$ and $0.75$, the energy difference $\Delta E_{c/a}$ between the tetragonal regular L1$_0$-type Heusler phase and the most favorable stable cubic structure lies in the range of thermal energies (cf.\ Table~\ref{table-1}).
Following the empirical relation $T_{\rm m} \approx k_{\mathrm{B}} \Delta E_{c/a}$, which has been proven quite successful in predicting the martensitic transformation temperatures in Heusler alloys\cite{Siewert-2011},  one may be tempted to expect a structural transition around $T=650\,$K for $x=0.75$ and around room temperature for $x=0.5$. However, as the finite temperature analysis in terms of the free energies in Sec.\ \ref{sec:finite} will show, contributions from mixing entropy are significant which can effectively inhibit such a transition and renders the above relation useless for the prediction of a diffusive transition.

\begin{table*}[t]
    \caption{Calculated total magnetic moment ($\mu_{tot}$ in $\mu_B$/f.u.), saturation magnetization~$M_s$~(in Am$^2$/kg), and MAE (in meV/f.u.\ and MJ/m$^3$) for the most stable structures (austenite and martensite) of Fe$_2$Ni$_{1+x}$Al$_{1-x}$. 
    For FeNi, the magnetic properties are also presented  for the T$^c$ structure. 
    Literature MAE values for L$1_0$-FeNi are mainly collected from theoretical studies using different methods (FLAPW-GGA~\cite{Wu-1999}, VASP-GGA~\cite{Miura-2013}, WIEN2k-GGA~\cite{Miura-2013,Edstrom-2014}, ASA-SPR-KKR-GGA~\cite{Edstrom-2014} and FP-SPR-KKR-GGA~\cite{Werwinski-2017}); the experimental ones marked with an asterisk (Refs.~\onlinecite{Pauleve-1962,Pauleve-1968, Poirier-2015,Lewis-2014}).
    }
    \centering
    \begin{ruledtabular}
    \begin{tabular}{llcccc cccll}
        \multirow{2}{*}{Composition} & \multicolumn{5}{c}{Austenite} & \multicolumn{5}{c}{Martensite}   \\ \cline{2-6} \cline{7-11}
         & \multirow{2}{*}{Structure}  & $\mu_{tot}$  & $M_s$ & \multicolumn{2}{c}{MAE}  & \multirow{2}{*}{Structure} &$\mu_{tot}$  & $M_s$& \multicolumn{2}{c}{MAE}   \\ 
            &   &   [$\mu_B$/f.u.]  &  [Am$^2$/kg] &[meV/f.u.]  & [MJ/m$^3$]    & &     [$\mu_B$/f.u.] & [Am$^2$/kg] & [meV/f.u.]  &  [MJ/m$^3$]     \\ 
        \hline
        
        $x = 0.0$   &   T$^p$   &   4.706  & 133.18  & 0.3075    &   1.0467     &   --- &   &    &    \\
        $x = 0.25$  &       T$^p$       &  4.856  & 132.10 & 0.0425   &  0.1468     & ---  &   &    &       \\
        $x = 0.50$  &    Inverse      & 5.617   & 147.13 & 0.0  &  0.0    & L$1_0$  & 5.722 & 149.87 &0.12  & 0.415      \\
        $x = 0.75$  & Inverse   &   6.044   & 152.65 &0.0  &   0.0  & L$1_0$  &   6.11 &  154.42 &0.21  & 0.7366  \\
        $x = 1.0$   &   --- &   &   &   &   &   L$1_0$ &   3.248 &  158.39 &0.0825  & 0.5867      \\ 
           &    &   &   &   &   &   T$^c$ &   3.255 &  158.72 &-0.0487  & -0.3463      \\ 
        &   &   &   &   &   &   &  &  &0.032 \cite{Wu-1999} &   0.22~\cite{Wu-1999}  \\ 
            &   &   &   &   &   &   &  & &0.078 \cite{Miura-2013}  &   0.56 \cite{Miura-2013}       \\ 
            &   &   &   &   &   &   &  & &0.069 \cite{Miura-2013,Edstrom-2014}  &   0.48 \cite{Miura-2013,Edstrom-2014}    \\ 
          &   &   &   &   &   &   &  & &0.11 \cite{Edstrom-2014}  &   0.77 \cite{Edstrom-2014}    \\ 
           &   &   &   &   &   &   &  & &0.032 \cite{Werwinski-2017}  &   0.22 \cite{Werwinski-2017}    \\          
                                &     &   &      &      &  &    &   &   & & 0.32 - 1.3$^{*}$   \\ 
    \end{tabular}
    \end{ruledtabular}
    \label{table-2}
\end{table*}

\subsection{Formation energies}\label{sec:formation}
 
We obtain insight into the structural stability of the compounds in the austenitic phase from the formation and decomposition energies of the fully optimized configurations with respect to atomic positions and cell parameters for each composition around $c/a = 1$.
In a first step, we evaluate  the~formation energy ($E_{\mathrm{form}}$) in terms of the difference of the total energy of the ternary compound and the weighted sum of the energies of the stable phases of the corresponding pure elements. This quantity is relevant e.\,g.\ for the preparation of thin films from the sputtering of elemental targets.
A negative sign of $E_{\mathrm{form}}$ 
points out the a stability of a compound under study against decomposition into its pure constituents.

This analysis reveals that the inverse structure at $c/a = 1$ keeps the cubic symmetry for all compositions, while T$^p$ and T$^\#$ structure remain almost cubic ($c/a \approx 0.98$ or 0.99) for compositions with $x = 0$ and 0.25. 
The T$^c$ structure also turns to be slightly tetragonally distorted  with $c/a$ about 1.036 and 1.064 for $x = 0$ and~0.25.
With further increase in Ni excess, T$^p$, T$^\#$, and T$^c$ structures deviate from the nearly cubic lattice vectors, which become clearly tetragonal.
According to the large energy difference with respect to other structures, we skip here the discussion of the regular L2$_1$ ordering. 
The ground state structure, lattice constants, and formation energy found for all compositions in austenite and martensite are listed in Table~\ref{table-1}. 
It~should be noted that in the case of $x=1$, the cubic structure with inverse atomic occupation is unstable due to a positive sign of~$E_{\mathrm{form}}$. 
 
Regarding the proper design of annealing procedures used to establish chemical order and long-term stability of the compounds, one may also be interested in the energy of decomposition of the compound into possible combinations with stable binary phases, which we will refer to as $E_{\mathrm{dec}}$. 
In~the following, we discuss as a particularly relevant case the relative stability of Fe$_2$NiAl against segregation into a dual phase-composite according to the equation:

\begin{equation}
%
 E_{\mathrm{dec}} = E_{\mathrm{Fe_2NiAl}} - \big[2E_{\mathrm{Fe}} + E_{\mathrm{NiAl}}\big],
\label{Eq2}
%
\end{equation}
where $E_{\mathrm{Fe_2NiAl}}$, $E_{\mathrm{Fe}}$, and $E_{\mathrm{NiAl}}$  are the total energies of Fe$_2$NiAl, {\it bcc} Fe and {\it bcc} NiAl, respectively. 
According to Eq.~(\ref{Eq2}), a negative value of $E_{\mathrm{dec}}$ indicates again the stability of a compound against a spinodal decomposition.

\begin{figure}[b] 
    \centering
    \includegraphics[width=0.9\columnwidth]{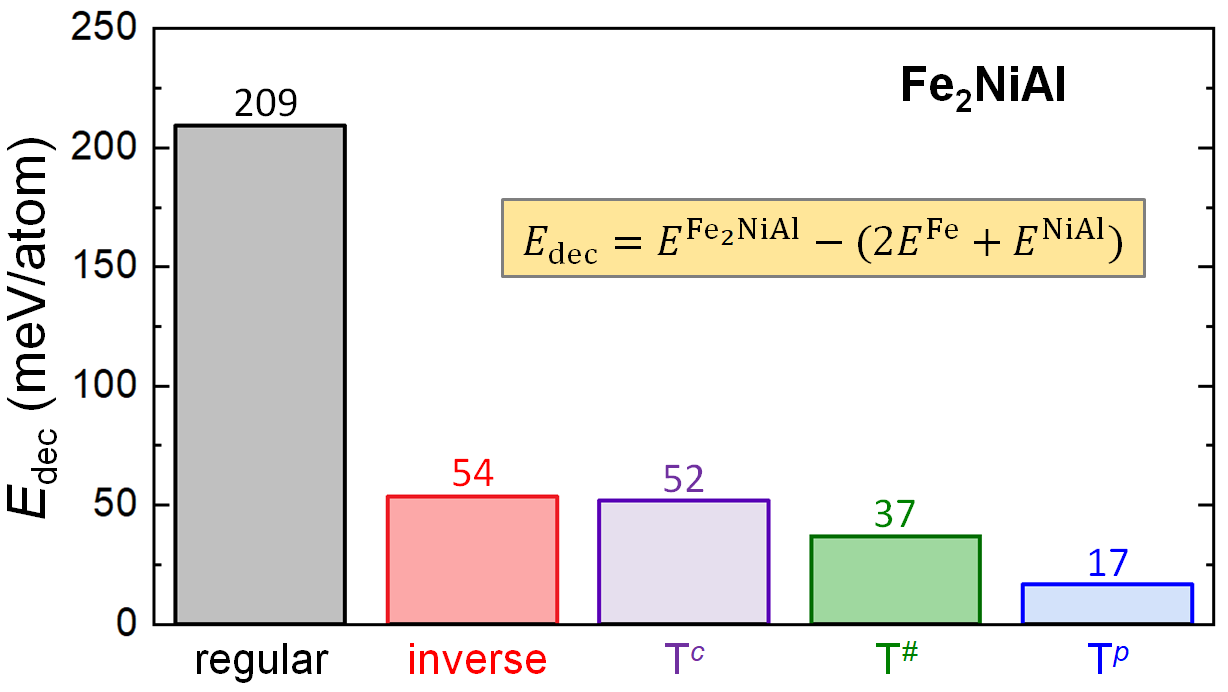} 
    \caption{
        Decomposition energy $E_{\mathrm{dec}}$ for the stoichiometric Fe$_2$NiAl compound with different crystalline structures against the segregation into a dual phase mixture of Fe and NiAl. 
           }
    \label{Fig-3}
    \end{figure}

Figure~\ref{Fig-3} illustrates the calculated $E_{\mathrm{dec}}$ for all considered structures of Fe$_2$NiAl. $E_{\mathrm{dec}}$ turns out to be positive for all  structures under investigation, which implies a long-term instability with respect to the above decomposition reaction. Nevertheless, the minimum value of $E_{\mathrm{dec}}=12.14\,$meV/atom found for Fe$_2$NiAl with the T$^p$ structure is rather small, corresponding to a temperature of around 100\,K. As one can see from Fig.~\ref{Fig-3}, this energy difference is about ten times smaller than for the regular and two times smaller than for the inverse Heusler structure, respectively, meaning that the layered T$^p$ structure is significantly less prone to a decomposition into pure Fe and binary NiAl systems as compared to the other structures.  
In fact, we verified  that T$^p$-Fe$_2$NiAl is stable in terms of its total energy against decomposition into the most relevant of binary and ternary compounds~-- except for the Eq.~(\ref{Eq2}); an comprehensive evaluation of the relative stability of T$^p$, inverse and conventional Fe$_2$NiAl is presented in the~SM.\cite{SM}
Keeping in mind that decomposition processes are significantly delayed at low temperatures and such small energies may be overcome by entropic contributions (which will be further discussed in Sec.~\ref{sec:finite}), one may be optimistic that with an appropriate scenario of preparation, Fe$_2$NiAl might be synthesized in a T$^p$ ordered structure and be sufficiently stable for application purposes.


\subsection{Magnetic moments and magnetocrystalline anisotropy}

We now turn to the magnetic properties of Fe$_2$Ni$_{1+x}$Al$_{1-x}$ in the lowest energy austenitic and martensitic structures. 
In Table~\ref{table-2}, we~present a summary of the total magnetic moment, saturation magnetization, and the MAE for the most preferred ordering motifs. 
Of particular interest for a potential application as a permanent magnet is in first line -- besides the total magnetic moment, which is sufficiently large in all cases -- the MAE as it determines the intrinsic coercivity of the material. We can note that we obtain positive MAE values for all relevant structures (except for T$^c$-FeNi) which corresponds to an uniaxial anisotropy as preferred in applications.
Our calculated MAE for L1$_0$-FeNi is comparable to previous theoretical studies and  consistent with experimental findings (see Table~\ref{table-2}). 
However, we predict the largest MAE of $\approx$ 0.3~meV/f.u. ($\approx 1.05$~MJ/m$^3$) for T$^p$-Fe$_2$NiAl. Despite the (almost) cubic lattice vectors,  
the alternating layers of Ni and Fe atoms result in a significant deviation from cubic symmetry (see Fig.~\ref{Fig-1}(e)), which allows for a significant uniaxial anisotropy. 
For~this compound, the MAE is about 2~times larger than that for the L1$_0$-FeNi. 
A hint for the origin of this increase can be obtained from the element-resolved contributions spin-orbit coupling~(SOC) term: For T$^p$-Fe$_2$NiAl, the absolute contributions from Fe and Ni turn out a factor of $2-3$ larger as compared to L$1_0$-FeNi, while the contributions from Ni are negative in both cases, diminishing the total MAE. This, however, becomes more relevant for L$1_0$-FeNi, where all non-spinpolarized Al atoms are replaced by Ni (for a complete discussion, see the SM\cite{SM}).

\begin{figure}[htp!!!] 
    \centering
    \includegraphics[width=0.35\textwidth]{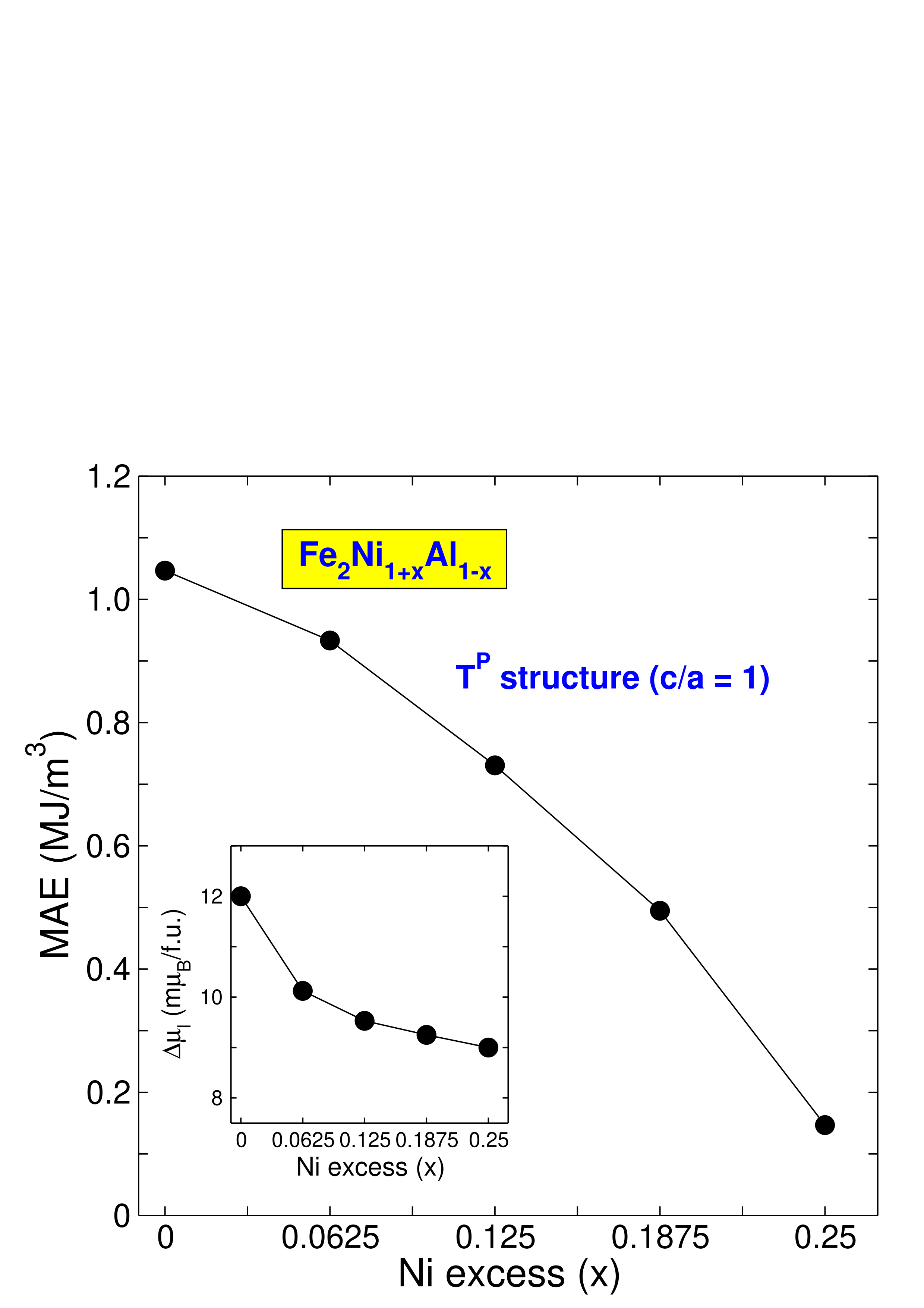}
    \caption{
        MAE of austenitic T$^p$ Fe$_2$Ni$_{1+x}$Al$_{1-x}$ (0 $\leq x \leq 0.25$) as a function of Ni excess. 
        The inset displays the orbital moment anisotropy $\Delta \mu_l$ behavior as a function of Ni content. 
           }
    \label{Fig-4}
    \end{figure}

As a consequence,
the replacement of Al by Ni excess atoms successively reduces the MAE in the austenitic phase and it vanishes when the inverse Heusler structure becomes favored, consistent with its cubic symmetry. In contrast, tetragonally distorted ternary L1$_0$-type Fe$_2$Ni$_{1+x}$Al$_{1-x}$  with $x = 0.5$ and 0.75, which represents the ground state in this composition range, demonstrates a considerable MAE being comparable of L1$_0$-FeNi with only little degradation.
To obtain a better resolution in the range of low Ni excess, we performed additional calculations for intermediate compositions with Ni excess between  $x = 0$ and 0.25 using a 64 atom supercell Fe$_{32}$Ni$_{16+n}$Al$_{16-n}$  with $n = 0$, 1, 2, 3, and 4, which corresponds to Fe$_2$Ni$_{1+x}$Al$_{1-x}$ with $x = 0$, 0.0625, 0.125, 0.1875, and 0.25, respectively. 
Figure~\ref{Fig-4} shows the corresponding dependence of the MAE with respect to the Ni excess concentration. Here we can confirm a smooth deviation of the MAE towards~$x = 0.25$.

Calculated MAE of L1$_0$-FeNi of about 0.59~MJ/m$^3$ is in a good agreement with the experimental data (Table~\ref{table-2}).
It~is worthwhile to calculate MAE of T$^c$-FeNi since the energy difference between the ground state L1$_0$ and the tetragonally distorted T$^c$ structure is with about 17~meV/atom comparatively small.
MAE~of T$^c$-FeNi is smaller in its absolute value and has moreover a negative sign, which tells us that the preferred orientation of the moments changes direction from the uniaxial easy axis to a cross-plane-direction. Thus, the competition of the different types of order might not only inhibit a perfect L1$_0$ ordering -- we must also expect a substantial impact regarding its hard-magnetic properties.

In addition to the MAE, we also calculated the orbital moment anisotropy, given by the difference $\Delta \mu_l=\mu_l^{001}-\mu_l^{100}$ between the orbital moments $\mu_l$ in the respective crystallographic directions, see Table~\ref{table-3}.
According to Patrick Bruno's relation derived from a perturbative treatment of MAE in thin ferromagnetic films of cubic metals~\cite{Bruno-1989}, the MAE is proportional to the difference between orbital magnetic moments oriented along easy and hard axis.
Hereby, the easy axis of the uniaxial anisotropy is parallel to the direction with the largest orbital moment, which is accounted for by the reverse order of the crystallographic directions in our definitions of the MAE and $\Delta\mu_l$.
Good agreement with this relation has been found in particular for elemental systems, but counter-examples have been reported for complex alloys and compounds (e.\,g., Refs.\ \onlinecite{Wilhelm-2001,Andersson-2007}). 
We see that -- except for the $L1_0$ martensite with $x=0.5$ -- for the relevant lowest energy structures indeed the easy axis coincides with the direction of the largest orbital moment.
Both, the uniaxial MAE and $\Delta \mu_l$, are maximal for stoichiometric composition and reduce with increasing Ni content. 
However, as demonstrated in the inset of Fig.~\ref{Fig-4}, the functional dependence is different indicating that the proportionality factor varies with composition. This indicates the presence of competing mechanisms connecting orbital moments and MAE, which are for instance well known from the related alloy FePt, where the proportionality factor even changes sign \cite{Solovyev-1995,Ravindran-2001,Gruner-2013}. 

In a recent high precision first-principles investigation, the decrease of the MAE with increasing disorder in FePt could be related to the presence of a rather low concentration of anti-site defects \cite{Wolloch-2017}. 
One may expect that also in the case of FeNi, the local environment contributes substantially to the rather complex dependence of MAE on composition and type of chemical order, since as in the case of FePt, the moment of the Ni-group element exhibits a predominantly itinerant (induced) character and is determined by the surrounding Fe moments, which are rather localized. 
Such itinerant character of Ni and a localized behavior of Mn is also a characteristic feature of Ni in Ni-Mn-based Heusler alloys \cite{Enkovaara-2003}.

\begin{table}[]
\caption{Total orbital magnetic moments $\mu_l$ calculated in [001] and [100] direction and their difference $\Delta \mu_l = \mu_l^{[001]} - \mu_l^{[100]}$ (in $\mu_{B}/f.u.$) for favorable structures of austenite and martensite Fe$_2$Ni$_{1+x}$Al$_{1-x}$. }
\begin{ruledtabular}
\begin{tabular}{lllll}
   Composition&  Structure  &   $\mu_l^{001}$  & $\mu_l^{100}$   & $\Delta \mu_l $  \\ \hline
\multicolumn{5}{c}{Austenite} \\ \hline
$x = 0.0$    &  T$^p$  & 0.113   & 0.101  & 0.012  \\
$x = 0.25$  &  T$^p$  &  0.123  &  0.114 &  0.009 \\
$x = 0.5$   &  Inverse  &  0.151  & 0.151  & 0.0  \\
$x = 0.75$  &  Inverse  &  0.166  & 0.166  & 0.0  \\ \hline
\multicolumn{5}{c}{Martensite} \\ \hline
$x = 0.5$    &  L1$_0$   &  0.155  & 0.157  & -0.002  \\
$x = 0.75$   &  L1$_0$   &  0.175  & 0.164  &  0.011 \\
$x = 1.0$    &  L1$_0$   &  0.177  & 0.170  &  0.007  \\ 
\end{tabular}
\end{ruledtabular}
\label{table-3}
\end{table}

\subsection{Exchange coupling constants}
\begin{figure}[htp!!!] 
    \centering
    \includegraphics[width=\columnwidth]{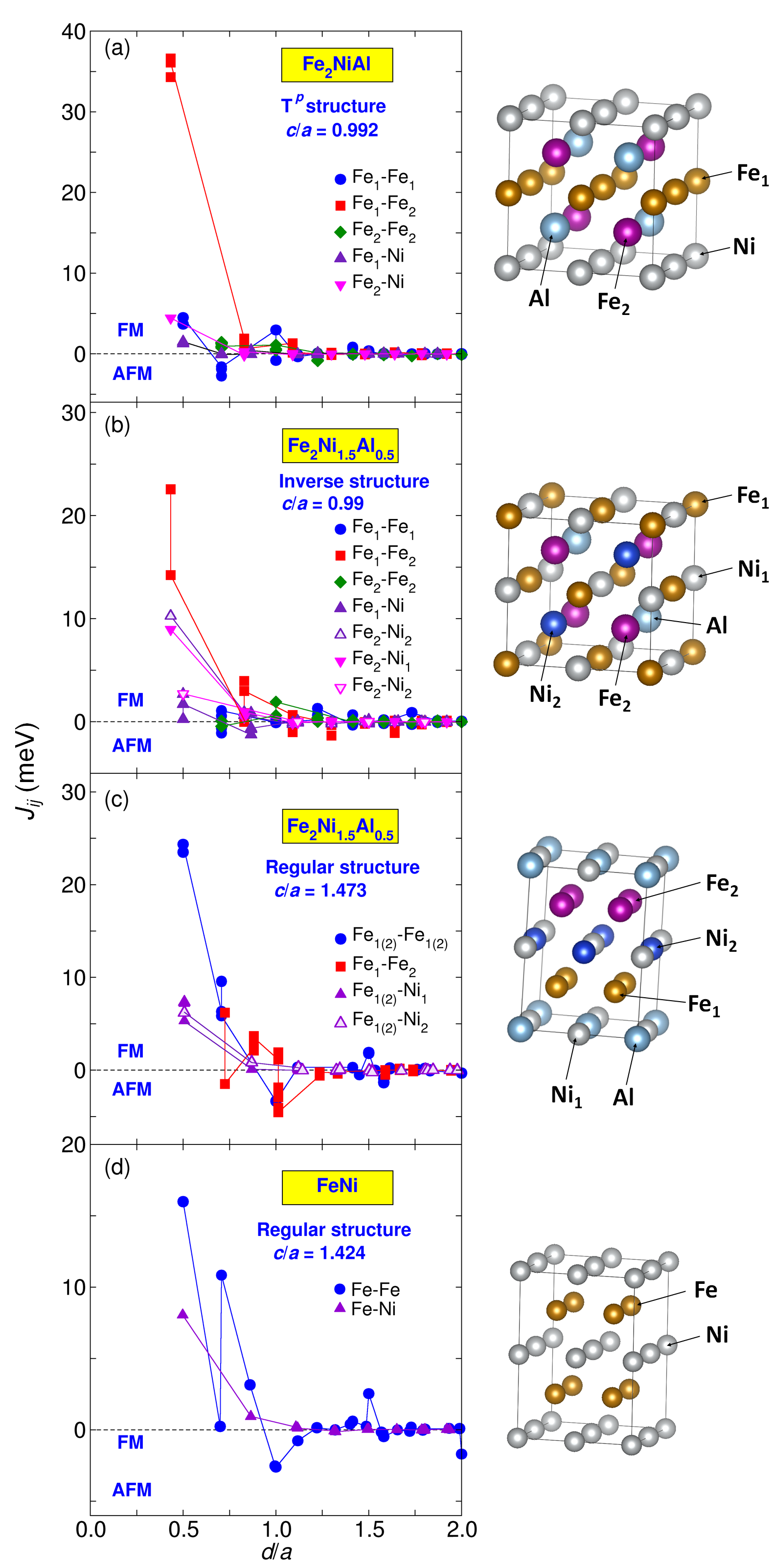}
    \caption{%
        Magnetic exchange parameters~$J_{ij}$ for the effective Heisenberg Hamiltonian between interacting transition metal atoms for Fe$_2$Ni$_{1+x}$Al$_{1-x}$ (0 $\leq x \leq 1$) with the most favorable crystalline structures in austenite and martensite, i.e., (a)~T$^p$ structure of stoichiometric Fe$_2$NiAl  with~$c/a=0.992$, 
        (b)~approximately cubic inverse Heusler structure for~$x=0.5$, 
        (c)~tetragonal regular L1$_0$ Heusler structure for $x=0.75$ and 
        (d)~tetratenite L1$_0$-FeNi. 
        The~exchange constants are depicted as a function of the interatomic distance~$d$ given in units of the respective lattice constant $a$ of the supercell. Positive values refer to ferromagnetic~(FM), negative values to antiferromagnetic~(AFM) pair interactions. 
        The~images on the right depict the corresponding supercell and inequivalent types of magnetic ions as used in the left column.
          }
    \label{Fig-5}
    \end{figure}

We have already seen that the magnetic properties of this compound are very sensitive to change in composition and atomic arrangement. 
In~this section, we investigate the effects of atomic arrangement in austenitic and martensitic phases of  Fe$_2$Ni$_{1+x}$Al$_{1-x}$ on the magnetic exchange interactions and Curie temperature in detail to unravel the impact of competing interatomic exchange interactions of Fe and Ni atoms at different sites, which are frequently observed in Heusler compounds. 
In~Fig.~\ref{Fig-5} we compare the evolution of the magnetic exchange constants $J_{ij}$ of the ground state structures in the investigated compositions range. The exchange constants define the interactions between different pairs of Fe and Ni atoms as a function of the distance $d$ between them in terms of a classical Heisenberg model Hamiltonian ${\cal H}_{\rm mag}=-\sum_{i\neq j}J_{ij}\,\vec{e}_i\cdot\vec{e}_j$, where $\vec{e}_i$  and $\vec{e}_j$ describe the unit vectors of the orientation of the magnetic spin moments at sites $i$ and $j$. 
The calculations were performed for the ground state structures obtained from the corresponding 16-atom supercell calculations. 
This implies two types of Fe and Ni atoms labeled as Fe$_1$, Fe$_2$ and Ni$_1$, Ni$_2$, which occupy structurally non-equivalent positions:  
In~case of the T$^p$ structure, four Fe$_1$ atoms lie in the same middle plane and four Fe$_2$ atoms occupy the alternating tetrahedral sites, see Fig.~\ref{Fig-1}(e).  
In~the case of inverse cubic structure, four Fe$_1$ and Fe$_2$ are placed at  $4a$(0,  0,  0)  and  $4c$(1/4,  1/4,  1/4)  sites, respectively, see Fig.~\ref{Fig-1}(b). 
Finally, for the tetragonal regular structure, four Fe$_1$ and Fe$_2$ atoms lie in the same bottom and top plane, correspondingly, see Fig.~\ref{Fig-1}(a). 
As~for Ni atoms, four Ni$_1$ are placed at regular Ni sites while Ni$_2$  are the excess Ni atoms, which substitute for Al.

As evident from the Figs.~\ref{Fig-5}(a,b), the intra-sublattice magnetic exchange constants reveal the largest FM coupling between nearest Fe$_1$-Fe$_2$ pairs of atoms located at the distance $d = (\sqrt{3}/4)\,a_0$ in the T$^p$ structure, which quickly approaches to zero from the third coordination sphere outwards. 
The inter-sublattice exchange couplings Fe$_{1(2)}$-Fe$_{1(2)}$ and Fe$_{1(2)}$-Ni$_{1(2)}$ are found to be smaller sufficiently and demonstrate a damped oscillating behavior with increasing distance. The~larger $\mu_{tot}$ for the composition with $x = 0.25$ can be related to the stronger FM coupling of Fe$_1$-Fe$_1$ and Fe$_1$-Ni$_2$ as compared to those for~$x=0$. 

In the case of alloys with inverse cubic structure, which is shown in Fig.~\ref{Fig-5}(b), the largest FM exchange interaction is observed between nearest Fe$_1$-Fe$_2$ pairs of atoms, similar to the T$^p$ structure. 
In~Fig.~\ref{Fig-5}(b) we observe a splitting of $J_{ij}$ into $\approx 22\,$meV and $\approx 14\,$meV between the Fe$_1$ and Fe$_2$ atoms located at $d = (\sqrt{3}/4)\,a_0$. The smaller exchange coupling constant is found for two Fe$_2$ atoms, which have Al neighbors at $d = 0.5\,a_0$.
The interactions Fe$_{1(2)}$-Fe$_{1(2)}$ and Fe$_{1(2)}$-Ni$_{1(2)}$ show again substantially smaller values except for Fe$_1$-Ni$_2$ and Fe$_2$-Ni$_1$, which are two times higher than those of T$^p$. 
This finding relates to the larger values of $\mu_{tot}$ for compositions with inverse structure in comparison to the T$^p$ structure (see Table~\ref{table-2}).

The~tetragonal distortion of the L1$_0$ martensite with a large $c/a$ ratio results in a clearly visible competition between FM and AFM exchange interactions of Fe pairs in the range of $0.7 < d/a < 1.2$ as shown in Fig.~\ref{Fig-5}(c). 
This is most pronounced for Fe$_1$ and Fe$_2$, which lie in the parallel planes of atoms that are spaced by a distance $c/2$~apart. 
Please note that in this case the Fe atoms in both layers (Fe$_1$ and Fe$_2$) are equivalent from the chemical point of view, but we keep the notation to visualize the difference between in-plane and out-of-plane interactions. 
The~nearest inter-sublattice Fe$_{1(2)}$-Fe$_{1(2)}$ pairs of atoms reveal the strongest FM exchange coupling, which is comparable to the nearest intra-sublattice Fe$_1$-Fe$_2$ interactions in the cubic inverse and T$^p$ structures, while cross-plane interactions are significantly reduced. 
The presence of a number of negative $J_{ij}$ could even lead to an effective decoupling of the layers, but the Fe-Ni interactions maintain the effective FM coupling between the layers mediated by the Ni atoms.
For the binary FeNi system shown in Fig.~\ref{Fig-5}(d), we observe again very similar situation: A considerably weaker exchange coupling between nearest Fe atoms in the plane, whereas the cross-plane interaction between next-nearest Fe almost vanishes. The~Fe-Ni interaction is slightly larger compared to the case  above and effectively strengthens the FM ordering of the Fe layers. 
In~both cases, we find hints for a damped oscillatory evolution of the Fe-Fe interactions with increasing~$d$.

From the full set of $J_{ij}$ we determined the Curie temperatures within the mean field approximation~($T_C^{\mathrm{MFA}}$), which are collected in Table~\ref{table-4} for the most relevant structures.  
For permanent magnet application of Heusler alloys, it is necessary to have both, a high Curie temperature and a large saturation magnetization. 
Indeed, we find that all considered compositions possess a rather high $T_C^{\mathrm{MFA}} > 1000$~K. 
As can be also seen from Table~\ref{table-4}, for both cubic inverse and tetragonal L$1_0$  structures, there is only a slight increase of $T_C^{\mathrm{MFA}}$ with increasing Ni content, since the different trends in intra- and inter-sublattice Fe-Fe and Fe-Ni interactions are effectively balanced out. 
The~largest value of $T_C^{\mathrm{MFA}}$ is obtained for the stoichiometric composition. 
The neglect of correlations close to the phase transition in the MFA as compared to a full statistical treatment in the framework of Monte Carlo simulations and the missing treatment of itinerant magnetism in localized spin models typically leads to an over-estimation of the Curie temperature, which easily reaches 20-30~\% of the experimental value.\cite{Garanin-1996,Sokolovskiy-2012,Meinert-2016,Wasilewski-2018,Wei-2018,Zagrebin-2020} Therefore we consider the calculated Curie temperature (and likewise also the saturation magnetization) of the Fe$_2$NiAl Heusler compound in proper agreement with the experimental value of 1010~K (and $M_{\rm s}=135\,$Am$^2$/kg)~\cite{Saito-2018}.
If~one replaces Al by Ni in the stoichiometric Fe$_2$NiAl in the T$^p$ arrangement, one observes a slight decrease in  $T_C^{\mathrm{MFA}}$ towards $x=0.25$, which results from the slight weakening of the nearest Fe$_1$-Fe$_2$ interaction and is much smaller compared to the rather substantial drop in the MAE in the same composition range. 

\begin{table}[]
\caption{Calculated Curie temperatures (in K) of austenite and martensite phases of the  Fe$_2$Ni$_{1+x}$Al$_{1-x}$ compounds using~MFA.  }
\begin{ruledtabular}
\begin{tabular}{llllll}
  Structure& $x=0.0$ &$x=0.25$  & $x=0.5$&$x=0.75$ &$x=1.0$ \\ \hline
T$^p$& 1387  &  1266  & ---   & ---  & ---  \\
Inverse& ---  &  ---   &  966  & 1024  & --- \\
L1$_0$&---   &  ---   &  1140  & 1159  & 1191  \\ 
\end{tabular}
\end{ruledtabular}
\label{table-4}
\end{table}

\subsection{Finite temperature calculations}\label{sec:finite}
As discussed in Sec.~\ref{sec:formation}, Fe-Ni-Al alloys near the stoichiometric composition are prone to a phase separation by spinodal decomposition, see also Refs. \onlinecite{Buschow-2007,Menushenkov-2015a,Menushenkov-2015b}. 
During the slow cooling from the high-temperature ordered {\it bcc} phase, Fe$_2$NiAl becomes unstable against the segregation and decomposes into two {\it bcc} phases consisting of FM Fe and a weak magnetic NiAl solid solution.
To optimize the magnetic properties, sophisticated heat treatment procedures, including various temperatures, waiting times, and also magnetic fields become important~\cite{Stanek-2010}. 
In~this section, we focus on the thermodynamic stability of the proposed structures in Fe$_2$Ni$_{1+x}$Al$_{1-x}$ by calculating the finite temperature free energy, 
\begin{equation}
    F = E_{\rm tot} + F_{\rm lat} + F_{\rm el}\,,
\end{equation}
which may help to develop further guidelines for the successful synthesis of the desired compound. Here, $E_{\rm tot}$ refers to the total energy for the respective structure from DFT calculations. $F_{\rm lat}$ represents the lattice free energy in the harmonic approximation derived from the VDOS of the respective structure calculated at the optimized lattice constant, which also includes the quantum mechanical zero point contribution to the lattice energy. Finally, $F_{\rm el}$ is the electronic free energy, which contains the change in internal energy $E_{\rm el}(T\!>\!0)$ at finite temperatures arising from the redistribution of electrons from occupied to unoccupied states according to the Fermi distribution function and the corresponding entropy $S_{\rm el}$. Both are estimated with sufficient accuracy within the Sommerfeld approximation, which relates $F_{\rm el}$ to the electronic density of states at the Fermi level $D(E_{\rm F})$:
\begin{equation}
F_{\rm el}(T)=E_{\rm el}(T\!>\!0) -T\,S_{\mathrm{el}}(T) \approx -\frac{\pi^2}{6}\,D(E_{\mathrm{F}})\,k_{\rm B}^2 T^2\,.
\end{equation}
The contribution $E_{\rm el}(T\!=\!0)$ is naturally included in $E_{\rm tot}$ obtained from our DFT calculations, 
 whereas we neglect in our work the finite temperature contributions to the free energy from the magnetic degrees of freedom.
 This can be considered an appropriate approximation at sufficiently low temperatures, where the magnetic moments can be considered in an almost perfectly ordered arrangement. In the vicinity of the Curie temperature, which is in our case in the order of 1000\,K and thus rather high, contributions from magnetic entropy may indeed become significant. In turn, the accurate prediction of the magnetic entropy at low temperatures in terms of a classical spin model is subject to difficulties which may arise, e.g., from the violation of Nernst's theorem in classical systems with continuous degrees of freedom. 
 The~solution requires advanced modeling approaches (see, e.\,g., Ref.~\onlinecite{Koermann-2010}), which is beyond the scope of the present investigation.

\begin{figure}[t] 
    \centering
    \includegraphics[width=\columnwidth]{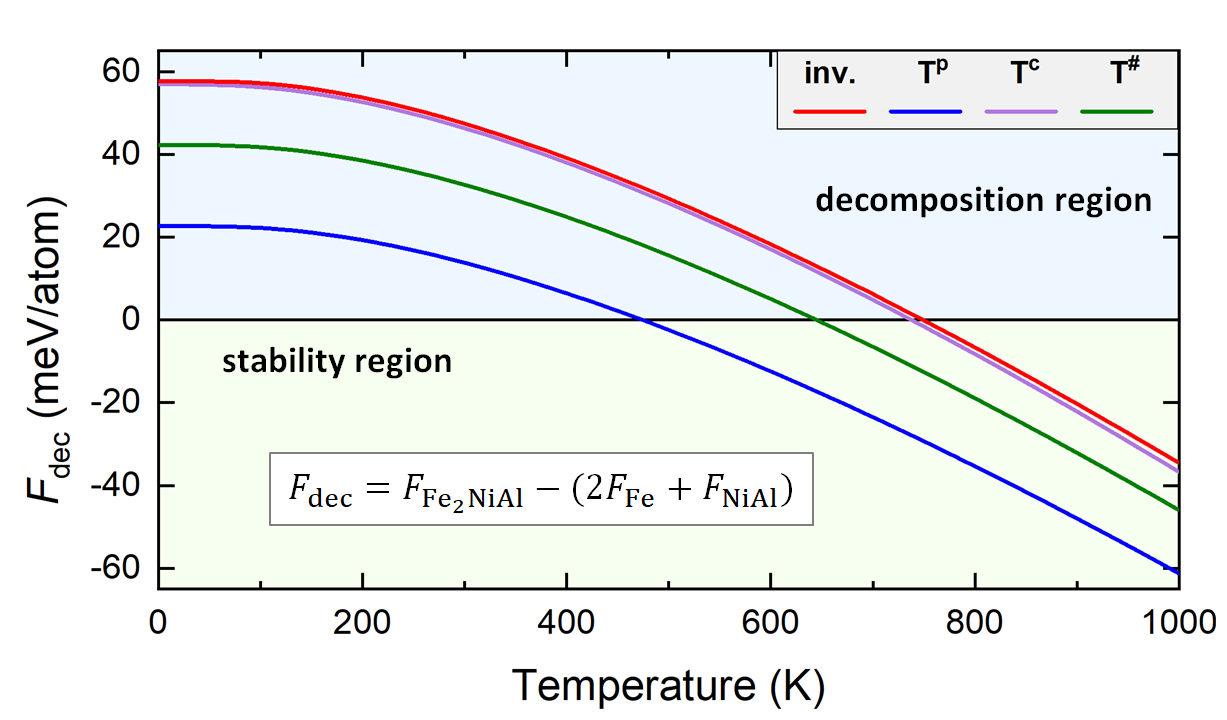}
    \caption{
        Decomposition of Fe$_2$NiAl into the pure Fe and binary NiAl in terms of the respective free energy difference as a function of temperature. 
   }
    \label{Fig_6}
    \end{figure}

Referring to Fig.\ \ref{Fig-3} of Sec.~\ref{sec:formation}, we will start our discussion once again from the stoichiometric case.
Figure~\ref{Fig_6} compares the stability of the four closely competing low energy structures of Fe$_2$NiAl, i.\,e., inverse Heusler, T$^c$, T$^\#$, and T$^p$ with respect to decomposition on the pure Fe and binary NiAl at temperatures up to $T=1000\,$K, in terms of the free energy difference $F_{\rm dec}=F_{\mathrm{Fe_2NiAl}} - ( 2\,F_{\mathrm{Fe}} + F_{\mathrm{NiAl}})$. Including the contributions from zero point energy, all structures retain their energetic order derived from $E_{\rm tot}$ 
which was shown in Fig.\ \ref{Fig-3} and remain susceptible to the above mentioned decomposition reaction. 
However, the finite temperature contributions to the free energy inhibit this process: The structure of Fe$_2$NiAl with the lowest energy, T$^p$, becomes stable against decomposition at $T=474\,$K, followed by T$^\#$ at $T=643\,$K, whereas T$^c$ and inverse Heusler  are predicted to be stable above 739 and 750~K, respectively. 
The~relative stability of the ternary structures with respect to each other does not change with temperature, leaving T$^p$ the most probable candidate for a fully ordered structure to be stabilized by an appropriate thermodynamic procedure.

Since T$^p$ corresponds to a free energy minimum in the range of realistic annealing temperatures, we can now discuss the stability of the off-stoichiometric Ni-excess compounds Fe$_2$Ni$_{1+x}$Al$_{1-x}$ with respect to a decomposition into the stoichiometric ternary compound T$^p$-Fe$_2$NiAl and binary~L1$_0$-FeNi. 
This bears some analogy with experimental findings of a decomposition tendency in Mn-excess Ni$_2$Mn$_{1+x}Z_{1-x}$ Heusler alloys, for which 
a dual-phase system consisting of the L2$_1$-cubic Ni$_2$Mn$Z$ phase and the L1$_0$-tetragonal NiMn phase has been reported~\cite{Yuhasz-2009,Yuhasz-2010,Krenke-2016,cakir-2016,Cakir-2017}. 
In order to fine-tune the temper-annealing conditions in the processing of the ternary system, it is important to have knowledge of the relative stability of particular ordering motifs at a given temperature in terms of the mixing free energy $F_{\mathrm{mix}}$:

\begin{eqnarray}
%
\label{eq:Fmix}
F_{\mathrm{mix}} & = &  
F^{\mathrm{aust/mart}}_{\mathrm{Fe_2Ni_{1+\textit{x}}Al_{1-\textit{x}}}} -T\,S_{\mathrm{mix}} \nonumber \\
& - & \big[ (1-x)F_{\mathrm{Fe_2NiAl}}^{T^p}  + x F_{\mathrm{FeNi}}^{L1_0} \big]\,, 
%
\end{eqnarray}
where $F^{\mathrm{aust/mart}}_{\mathrm{Fe_2Ni_{1+\textit{x}}Al_{1-\textit{x}}}}$ is the free energy of the ternary, off-stoichiometric compositions under consideration in austenitic or martensitic phase; $F_{\mathrm{Fe_2NiAl}}^{T^p}$ and $F_{\mathrm{FeNi}}^{L1_0}$ are the corresponding free energies of Fe$_2$NiAl with the T$^p$ cubic structure and FeNi with the L1$_0$ tetragonal structure; $S_{\mathrm{mix}}$ accounts the mixing entropy arising from the partial disorder in the off-stoichiometric ternary system, where the excess Ni is distributed randomly over the four sublattices:

\begin{equation}
\label{eq:Smix}
S_{\mathrm{mix}}=-\frac{1}{4}k_{\rm B} \sum \limits_{i=1}^4\big[  x_i  \ln x_i + (1-x_i) \ln (1-x_i)\big]\,.
\end{equation}
Here $x_i$ is the concentration of Ni on the $i^{\rm th}$ sublattice and $k_{\rm B}$ is the Boltzmann constant.
Since these structures are fully ordered, $S_{\mathrm{mix}}=0$ for stoichiometric Fe$_2$NiAl and FeNi and is thus left out in Eq.~(\ref{eq:Fmix}).

\begin{figure}[b] 
    \centering
    \includegraphics[width=\columnwidth]{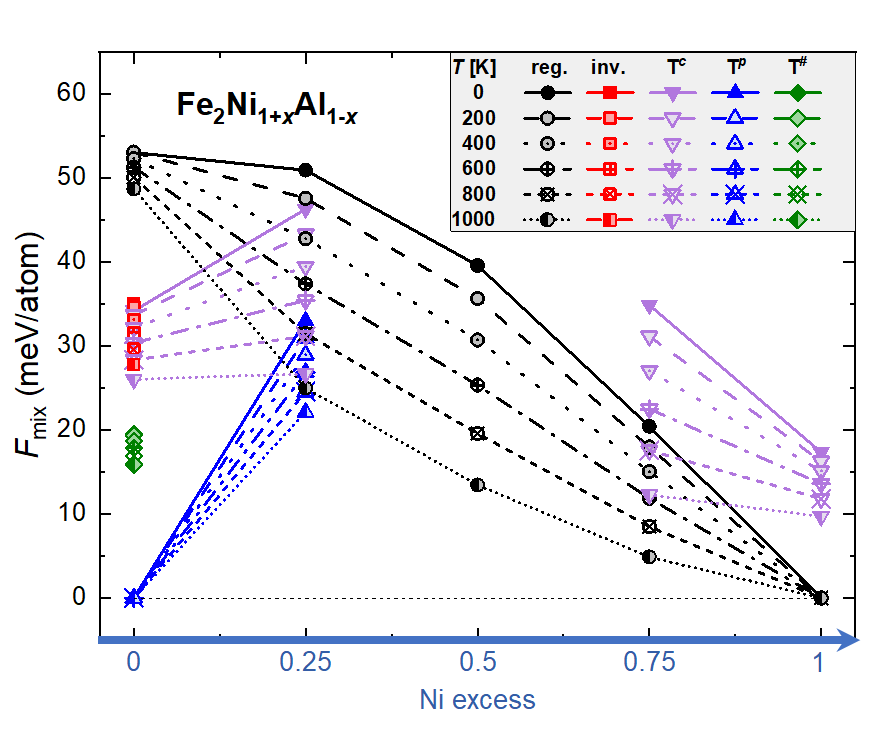}
    \caption{
        The dependence of mixing free energy $F_{\rm mix}$ on the Ni-excess concentration
        for selected FM austenite and martensite structures of Fe$_2$Ni$_{1+x}$Al$_{1-x}$
        at various temperatures in the range between 0 and 1000~K.
        }
    \label{Fig-7}
    \end{figure}

The resulting behavior of the mixing free energy at different temperatures as a function of composition is depicted in Fig.~\ref{Fig-7}.
For $T=0$, it omits all finite temperature contributions except for the zero point energy.
For~stoichiometric Fe$_2$NiAl, where mixing entropy vanishes for all compounds under consideration, T$^\#$ is in accordance with Fig.\ \ref{Fig-3} the closest competitor to the energetically most favorable T$^p$ structure, which becomes closer in energy with increasing temperature. On the other side, regular L1$_0$-FeNi proves preferable to the $T^c$ structure with all Al replaced by Ni in the range of realistic temperatures.
%

For the non-stoichiometric compositions, the impact of temperature on $F_{\rm mix}$ increases substantially due to the presence of the mixing entropy. This leads in particular for the regular L1$_0$ to a significant reduction of the difference in free energy compared to a respective phase mixture of the stoichiometric compounds, which may stabilize L1$_0$-type solid solutions at the very Ni-rich content at high temperatures. 
Still, the off-stoichiometic compounds remain unstable in general against this decomposition for all temperatures under consideration. A very similar trend was reported recently for off-stoichiometric Ni-Mn-based Heusler compounds \cite{Entel-2018,Sokolovskiy-2019}. 
For the composition with~$x=0.25$, we see a rather close competition between T$^c$ and T$^p$ (both with nearly cubic lattice parameters) and L1$_0$ martensite. The inverse Heusler structure was not considered here, since according to Fig.~\ref{Fig-2} its energy minimum is clearly above the other candidates. 
In~turn, we observe that at $x=0.75$ the T$^c$ phase with tetragonal lattice parameters becomes with increasing temperatures increasingly competitive with the martensitic L1$_0$ phase.

\begin{figure}[t] 
    \centering
    \includegraphics[width=\columnwidth]{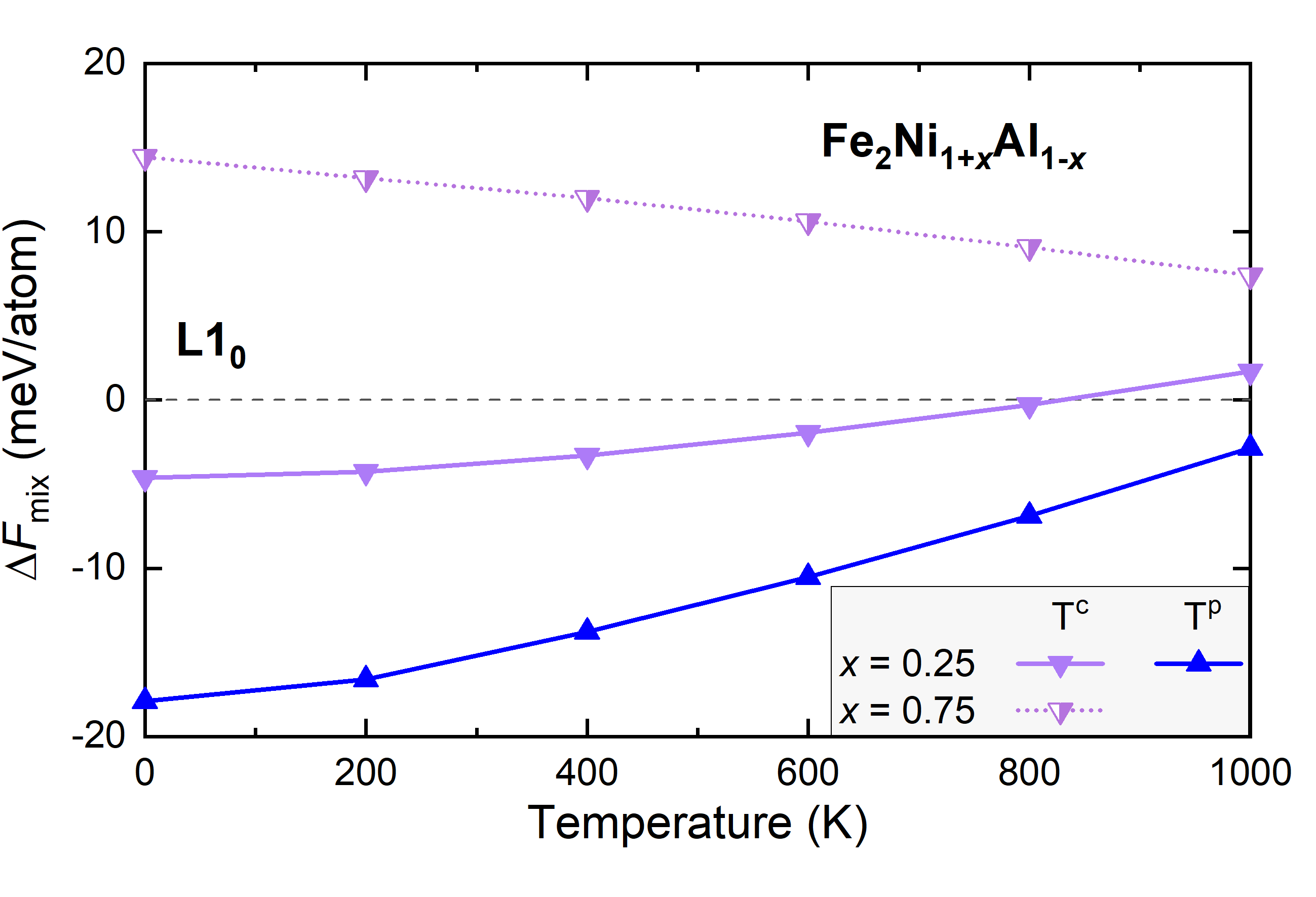}
    \caption{
        Temperature variation of the free energy of the austenitic  T$^p$ and T$^c$ phase at $x=0.25$ and martensitic T$^c$ at $x=0.75$. The free energy is plotted as the difference 
        $\Delta F_{\rm mix}$ to the free energy of the regular L1$_0$ structure at the respective composition.
        }
    \label{Fig-8}
    \end{figure}

In Fig.~\ref{Fig-8} we show the temperature dependence of the free energy of T$^p$ (at $x=0.25$) and T$^c$ ($x=0.25$ and $x=0.75$) both at the respective optimum lattice parameters relative to the free energy of the tetragonal L1$_0$ phase with the respective composition, i.\,e., 
$\Delta F_{\rm mix}=F_{\rm mix}^{{\rm T}^p,{\rm T}^c}-F_{\rm mix}^{{\rm L1}_0}$. 
We see that for $x=0.25$ the T$^c$ structure is more stable than L1$_0$ for temperatures below $T=800\,$K, 
while T$^p$ preserves the lowest free energy in the entire temperature range. 
However, extrapolation of $\Delta F_{\rm mix}$ implies that a crossover may occur somewhere below $T=1200\,$K and 
the temperature of the crossing point may decrease further, if a larger amount of Al is substituted by~Ni. 
On the Ni-rich side only T$^c$ is competitive but its free energy remains considerably above the L1$_0$ phase. 
Although here $\Delta F_{\rm mix}$ decreases with temperature, we can not predict a stabilization at a reasonable temperature. 
However, since the sign and the slope of $\Delta F_{\rm mix}$ changes between $x=0.25$ and $x=0.75$ we might rather expect interesting cross-over effects between these structures at intermediate compositions. 
To resolve the corresponding phase transitions from first principles with sufficient accuracy, better structural models for the partially disordered phases are needed, which require significantly larger supercells. 
As~this severely increases the computational demands, the corresponding phonon calculations have been left for future investigations.

%

\section{Conclusions}

The substitutional series of the Heusler compounds Fe$_2$Ni$_{1+x}$Al$_{1-x}$ ($0 \leq x \leq 1$) in austenitic and martensitic phases with different kinds of atomic arrangements were investigated theoretically to address their phase stability as well as structural and magnetic properties including the MAE. The present study was motivated by the potential importance of Fe$_2$NiAl and L1$_0$-FeNi for a permanent magnet application. It predicts phase segregation tendencies in Fe$_2$NiAl and addresses the complexity of the synthesis of FeNi with L1$_0$ order which may arise in part from the presence of competing structures. 
We~learn how the phase stability and MAE can be controlled by exchanging the positions of Fe and Ni atoms, by considering three kinds of atomic arrangements which can be derived from the inverse Heusler structure of Fe$_2$NiAl by shuffling Fe and Ni atoms on particular sites.
To~this end, the formation and mixing energies of a series of compositions with considered crystalline structures were calculated at zero and finite-temperatures. Our calculations of the saturation magnetization, exchange constants, and finally Curie temperature confirm that all considered compounds exhibit sufficiently large values, as required for permanent magnet applications.

For the austenitic phase of Fe$_2$NiAl, we found that all recently proposed structures denoted as T$^c$, T$^\#$ and T$^p$ are lower in energy compared to the inverse and regular structures, which were mainly discussed so far in the literature. 
The~ground state corresponds to the ordered T$^p$ structure, which was proposed recently as a stable low temperature structure for some quarternary stoichiometric Heusler alloys \cite{Neibecker-2017}. 
It~consists of alternating layers of pure Fe and pure Ni with mixed layers of Fe and Al in between and {\it bcc}-type coordination of the ions.
This structure was also found to possess the largest uniaxial MAE in the study, 
due to the intrinsic non-cubic arrangement of chemical elements in the cell. The MAE reaches a value of about 1.05~MJ/m$^3$, exceeding the MAE of L1$_0$-FeNi.
This is accompanied in the case of T$^p$-Fe$_2$NiAl by a larger anisotropy of the orbital magnetic moments. 
In the T$^p$ structure, the~partial substitution of Al by Ni yields a successive reduction of the MAE, which becomes nearly suppressed for Fe$_2$Ni$_{1.25}$Al$_{0.75}$.
Further increase in the off-stoichiometry leads to a crossover of the structural motifs and at large Ni content finally to the appearance of tetragonal L1$_0$ order.
In~the case of binary FeNi, the T$^c$ structure with a staggered arrangement of Fe and Ni turns out to be rather close in energy to the L$1_0$ phase. T$^c$-FeNi possesses a planar MAE, with an absolute value only about half as large as the one of L$1_0$-FeNi and the opposite sign.     

Regarding the decomposition of Fe$_2$NiAl into {\it bcc}-Fe and NiAl, our total energy calculations reveal that the T$^p$ structure has a positive formation energy ($E_{\mathrm{form}}~\approx~12.4$~meV/atom) at $T = 0$~K being several times lower than other structures including the regular and inverse Heusler structure.  
Despite still lying above the convex hull, it can be expected that T$^p$-Fe$_2$NiAl can be stabilized by a proper choice of annealing condition, since due to the contributions from vibrational entropy calculated in the harmonic approximation, the free energy of the T$^p$ phase becomes lower compared to a phase mixture of Fe and NiAl above $T=474\,$K, which is reasonable for typical temper-annealing procedures.
The finite temperature approach to study the segregation tendency in Ni-rich Fe$_2$Ni$_{1+x}$Al$_{1-x}$ have shown that in particular the contributions from mixing entropy render the free energy quite sensitive to temperature. This reduces the free energy offset to a phase mixture between stoichiometric Fe$_2$NiAl and FeNi substantially, which enhances the probability to obtain off-stoichiometric phases. Nevertheless, all off-stoichiometric compounds demonstrate the tendency to decompose into a dual-phase mixture consisting of Fe$_2$NiAl and FeNi at temperatures below $T=1000\,$K.
 
In~general, we consider the fundamental understanding of the role of atomic arrangement in a crystalline structure in Heusler alloys of great interest for both the shape memory and the permanent magnet community as this might be employed to deliberately modify the phase stability, structural and magnetic properties as well as~MAE.
We~believe that the T$^p$ structure in Fe$_2$NiAl and related materials could be interesting and promising candidate for systems exhibiting a large intrinsic MAE, saturation magnetization and high Curie temperature, suitable for a potential application in low-cost hard magnets.

\section{Acknowledgments}
The authors dedicate this work to Prof.~Dr.~Peter Entel, who passed away on Aug.~4,~2021. 
We~are grateful for the involved discussions, his brilliant ideas and substantial support which fostered many successful common projects over the past decades, including this one.
This~work was further supported by the RSF~-~Russian Science Foundation project No.~17-72-20022 (geometric optimization and mixing energy and Exchange parameters calculations).
MAE~calculations were performed with the support of the Ministry of Science and Higher Education of the Russian Federation within the framework of the Russian State Assignment under contract No.~075-00250-20-03.
Phonon and mixing energy calculations were performed on the MagnitUDE high performance computing system of the University of Duisburg-Essen (DFG INST 20876/209-1 and 20876/243-1 FUGG) with financial support from the Deutsche Forschungsgemeinschaft (DFG, German Research Foundation) within TRR~270 (subproject~B06), Project-ID 405553726.

\bibliography{liter}

\begin{thebibliography}{91}
\expandafter\ifx\csname natexlab\endcsname\relax\def\natexlab#1{#1}\fi
\expandafter\ifx\csname bibnamefont\endcsname\relax
  \def\bibnamefont#1{#1}\fi
\expandafter\ifx\csname bibfnamefont\endcsname\relax
  \def\bibfnamefont#1{#1}\fi
\expandafter\ifx\csname citenamefont\endcsname\relax
  \def\citenamefont#1{#1}\fi
\expandafter\ifx\csname url\endcsname\relax
  \def\url#1{\texttt{#1}}\fi
\expandafter\ifx\csname urlprefix\endcsname\relax\def\urlprefix{URL }\fi
\providecommand{\bibinfo}[2]{#2}
\providecommand{\eprint}[2][]{\url{#2}}

\bibitem[{\citenamefont{Skomski and Coey}(2016)}]{Skomski-2016}
\bibinfo{author}{\bibfnamefont{R.}~\bibnamefont{Skomski}} \bibnamefont{and}
  \bibinfo{author}{\bibfnamefont{J.}~\bibnamefont{Coey}},
  \bibinfo{journal}{Scr. Mater.} \textbf{\bibinfo{volume}{112}},
  \bibinfo{pages}{3} (\bibinfo{year}{2016}).

\bibitem[{\citenamefont{Skomski}(2016)}]{Skomski-2016b}
\bibinfo{author}{\bibfnamefont{R.}~\bibnamefont{Skomski}},
  \emph{\bibinfo{title}{Permanent Magnets: History, Current Research, and
  Outlook}} (\bibinfo{publisher}{Springer International Publishing},
  \bibinfo{address}{Cham}, \bibinfo{year}{2016}), pp.
  \bibinfo{pages}{359--395}.

\bibitem[{\citenamefont{Hono and Sepehri-Amin}(2018)}]{Hono-2018}
\bibinfo{author}{\bibfnamefont{K.}~\bibnamefont{Hono}} \bibnamefont{and}
  \bibinfo{author}{\bibfnamefont{H.}~\bibnamefont{Sepehri-Amin}},
  \bibinfo{journal}{Scr. Mater.} \textbf{\bibinfo{volume}{151}},
  \bibinfo{pages}{6} (\bibinfo{year}{2018}).

\bibitem[{\citenamefont{Skokov and Gutfleisch}(2018)}]{Skokov-2018}
\bibinfo{author}{\bibfnamefont{K.}~\bibnamefont{Skokov}} \bibnamefont{and}
  \bibinfo{author}{\bibfnamefont{O.}~\bibnamefont{Gutfleisch}},
  \bibinfo{journal}{Scr. Mater.} \textbf{\bibinfo{volume}{154}},
  \bibinfo{pages}{289} (\bibinfo{year}{2018}).

\bibitem[{\citenamefont{Mohapatra and Liu}(2018)}]{Mohapatra-2018}
\bibinfo{author}{\bibfnamefont{J.}~\bibnamefont{Mohapatra}} \bibnamefont{and}
  \bibinfo{author}{\bibfnamefont{J.~P.} \bibnamefont{Liu}},
  \emph{\bibinfo{title}{Rare-Earth-free permanent magnets: the past and
  future}} (\bibinfo{publisher}{Elsevier}, \bibinfo{year}{2018}),
  vol.~\bibinfo{volume}{27}, pp. \bibinfo{pages}{1--57}.

\bibitem[{\citenamefont{Kovacs et~al.}(2020)\citenamefont{Kovacs, Fischbacher,
  Gusenbauer, Oezelt, Herper, Vekilova, Nieves, Arapan, and
  Schrefl}}]{Kovacs-2020}
\bibinfo{author}{\bibfnamefont{A.}~\bibnamefont{Kovacs}},
  \bibinfo{author}{\bibfnamefont{J.}~\bibnamefont{Fischbacher}},
  \bibinfo{author}{\bibfnamefont{M.}~\bibnamefont{Gusenbauer}},
  \bibinfo{author}{\bibfnamefont{H.}~\bibnamefont{Oezelt}},
  \bibinfo{author}{\bibfnamefont{H.~C.} \bibnamefont{Herper}},
  \bibinfo{author}{\bibfnamefont{O.~Y.} \bibnamefont{Vekilova}},
  \bibinfo{author}{\bibfnamefont{P.}~\bibnamefont{Nieves}},
  \bibinfo{author}{\bibfnamefont{S.}~\bibnamefont{Arapan}}, \bibnamefont{and}
  \bibinfo{author}{\bibfnamefont{T.}~\bibnamefont{Schrefl}},
  \bibinfo{journal}{Engineering} \textbf{\bibinfo{volume}{6}},
  \bibinfo{pages}{148} (\bibinfo{year}{2020}).

\bibitem[{\citenamefont{Coey}(2020)}]{Coey-2020}
\bibinfo{author}{\bibfnamefont{J.}~\bibnamefont{Coey}},
  \bibinfo{journal}{Engineering} \textbf{\bibinfo{volume}{6}},
  \bibinfo{pages}{119} (\bibinfo{year}{2020}).

\bibitem[{\citenamefont{Kuz’min et~al.}(2014)\citenamefont{Kuz’min, Skokov,
  Jian, Radulov, and Gutfleisch}}]{Kuzmin-2014}
\bibinfo{author}{\bibfnamefont{M.}~\bibnamefont{Kuz’min}},
  \bibinfo{author}{\bibfnamefont{K.}~\bibnamefont{Skokov}},
  \bibinfo{author}{\bibfnamefont{H.}~\bibnamefont{Jian}},
  \bibinfo{author}{\bibfnamefont{I.}~\bibnamefont{Radulov}}, \bibnamefont{and}
  \bibinfo{author}{\bibfnamefont{O.}~\bibnamefont{Gutfleisch}},
  \bibinfo{journal}{J. Phys. Condens. Matter.} \textbf{\bibinfo{volume}{26}},
  \bibinfo{pages}{064205} (\bibinfo{year}{2014}).

\bibitem[{\citenamefont{McCallum et~al.}(2014)\citenamefont{McCallum, Lewis,
  Skomski, Kramer, and Anderson}}]{Mccallum-2014}
\bibinfo{author}{\bibfnamefont{R.}~\bibnamefont{McCallum}},
  \bibinfo{author}{\bibfnamefont{L.~H.} \bibnamefont{Lewis}},
  \bibinfo{author}{\bibfnamefont{R.}~\bibnamefont{Skomski}},
  \bibinfo{author}{\bibfnamefont{M.}~\bibnamefont{Kramer}}, \bibnamefont{and}
  \bibinfo{author}{\bibfnamefont{I.}~\bibnamefont{Anderson}},
  \bibinfo{journal}{Annu. Rev. Mater. Res.} \textbf{\bibinfo{volume}{44}},
  \bibinfo{pages}{451} (\bibinfo{year}{2014}).

\bibitem[{\citenamefont{Niarchos et~al.}(2015)\citenamefont{Niarchos,
  Giannopoulos, Gjoka, Sarafidis, Psycharis, Rusz, Edstr{\"o}m, Eriksson,
  Toson, Fidler et~al.}}]{Niarchos-2015}
\bibinfo{author}{\bibfnamefont{D.}~\bibnamefont{Niarchos}},
  \bibinfo{author}{\bibfnamefont{G.}~\bibnamefont{Giannopoulos}},
  \bibinfo{author}{\bibfnamefont{M.}~\bibnamefont{Gjoka}},
  \bibinfo{author}{\bibfnamefont{C.}~\bibnamefont{Sarafidis}},
  \bibinfo{author}{\bibfnamefont{V.}~\bibnamefont{Psycharis}},
  \bibinfo{author}{\bibfnamefont{J.}~\bibnamefont{Rusz}},
  \bibinfo{author}{\bibfnamefont{A.}~\bibnamefont{Edstr{\"o}m}},
  \bibinfo{author}{\bibfnamefont{O.}~\bibnamefont{Eriksson}},
  \bibinfo{author}{\bibfnamefont{P.}~\bibnamefont{Toson}},
  \bibinfo{author}{\bibfnamefont{J.}~\bibnamefont{Fidler}},
  \bibnamefont{et~al.}, \bibinfo{journal}{JOM} \textbf{\bibinfo{volume}{67}},
  \bibinfo{pages}{1318} (\bibinfo{year}{2015}).

\bibitem[{\citenamefont{Hirosawa}(2015)}]{Hirosawa-2015}
\bibinfo{author}{\bibfnamefont{S.}~\bibnamefont{Hirosawa}},
  \bibinfo{journal}{J. Magn. Soc. Jpn.} \textbf{\bibinfo{volume}{39}},
  \bibinfo{pages}{85} (\bibinfo{year}{2015}).

\bibitem[{\citenamefont{Li et~al.}(2016)\citenamefont{Li, Pan, Li, and
  Zhang}}]{Li-2016}
\bibinfo{author}{\bibfnamefont{D.}~\bibnamefont{Li}},
  \bibinfo{author}{\bibfnamefont{D.}~\bibnamefont{Pan}},
  \bibinfo{author}{\bibfnamefont{S.}~\bibnamefont{Li}}, \bibnamefont{and}
  \bibinfo{author}{\bibfnamefont{Z.}~\bibnamefont{Zhang}},
  \bibinfo{journal}{Sci. China Phys. Mech. Astron.}
  \textbf{\bibinfo{volume}{59}}, \bibinfo{pages}{617501}
  (\bibinfo{year}{2016}).

\bibitem[{\citenamefont{Edstr{\"o}m et~al.}(2014)\citenamefont{Edstr{\"o}m,
  Chico, Jakobsson, Bergman, and Rusz}}]{Edstrom-2014}
\bibinfo{author}{\bibfnamefont{A.}~\bibnamefont{Edstr{\"o}m}},
  \bibinfo{author}{\bibfnamefont{J.}~\bibnamefont{Chico}},
  \bibinfo{author}{\bibfnamefont{A.}~\bibnamefont{Jakobsson}},
  \bibinfo{author}{\bibfnamefont{A.}~\bibnamefont{Bergman}}, \bibnamefont{and}
  \bibinfo{author}{\bibfnamefont{J.}~\bibnamefont{Rusz}},
  \bibinfo{journal}{Phys. Rev. B} \textbf{\bibinfo{volume}{90}},
  \bibinfo{pages}{014402} (\bibinfo{year}{2014}).

\bibitem[{\citenamefont{Werwi{\'n}ski and Marciniak}(2017)}]{Werwinski-2017}
\bibinfo{author}{\bibfnamefont{M.}~\bibnamefont{Werwi{\'n}ski}}
  \bibnamefont{and}
  \bibinfo{author}{\bibfnamefont{W.}~\bibnamefont{Marciniak}},
  \bibinfo{journal}{J. Phys. D: Appl. Phys.} \textbf{\bibinfo{volume}{50}},
  \bibinfo{pages}{495008} (\bibinfo{year}{2017}).

\bibitem[{\citenamefont{Cui et~al.}(2018)\citenamefont{Cui, Kramer, Zhou, Liu,
  Gabay, Hadjipanayis, Balasubramanian, and Sellmyer}}]{Cui-2018}
\bibinfo{author}{\bibfnamefont{J.}~\bibnamefont{Cui}},
  \bibinfo{author}{\bibfnamefont{M.}~\bibnamefont{Kramer}},
  \bibinfo{author}{\bibfnamefont{L.}~\bibnamefont{Zhou}},
  \bibinfo{author}{\bibfnamefont{F.}~\bibnamefont{Liu}},
  \bibinfo{author}{\bibfnamefont{A.}~\bibnamefont{Gabay}},
  \bibinfo{author}{\bibfnamefont{G.}~\bibnamefont{Hadjipanayis}},
  \bibinfo{author}{\bibfnamefont{B.}~\bibnamefont{Balasubramanian}},
  \bibnamefont{and} \bibinfo{author}{\bibfnamefont{D.}~\bibnamefont{Sellmyer}},
  \bibinfo{journal}{Acta Materialia} \textbf{\bibinfo{volume}{158}},
  \bibinfo{pages}{118} (\bibinfo{year}{2018}).

\bibitem[{\citenamefont{Tian et~al.}(2019)\citenamefont{Tian, Lev{\"a}m{\"a}ki,
  Eriksson, Kokko, Nagy, D{\'e}lczeg-Czirj{\'a}k, and Vitos}}]{Tian-2019}
\bibinfo{author}{\bibfnamefont{L.-Y.} \bibnamefont{Tian}},
  \bibinfo{author}{\bibfnamefont{H.}~\bibnamefont{Lev{\"a}m{\"a}ki}},
  \bibinfo{author}{\bibfnamefont{O.}~\bibnamefont{Eriksson}},
  \bibinfo{author}{\bibfnamefont{K.}~\bibnamefont{Kokko}},
  \bibinfo{author}{\bibfnamefont{{\'A}.}~\bibnamefont{Nagy}},
  \bibinfo{author}{\bibfnamefont{E.~K.} \bibnamefont{D{\'e}lczeg-Czirj{\'a}k}},
  \bibnamefont{and} \bibinfo{author}{\bibfnamefont{L.}~\bibnamefont{Vitos}},
  \bibinfo{journal}{Sci. Rep.} \textbf{\bibinfo{volume}{9}}, \bibinfo{pages}{1}
  (\bibinfo{year}{2019}).

\bibitem[{\citenamefont{Tian et~al.}(2020)\citenamefont{Tian, Eriksson, and
  Vitos}}]{Tian-2020}
\bibinfo{author}{\bibfnamefont{L.-Y.} \bibnamefont{Tian}},
  \bibinfo{author}{\bibfnamefont{O.}~\bibnamefont{Eriksson}}, \bibnamefont{and}
  \bibinfo{author}{\bibfnamefont{L.}~\bibnamefont{Vitos}},
  \bibinfo{journal}{Sci. Rep.} \textbf{\bibinfo{volume}{10}},
  \bibinfo{pages}{1} (\bibinfo{year}{2020}).

\bibitem[{\citenamefont{Tian et~al.}(2021)\citenamefont{Tian, Gutfleisch,
  Eriksson, and Vitos}}]{Tian-2021}
\bibinfo{author}{\bibfnamefont{L.-Y.} \bibnamefont{Tian}},
  \bibinfo{author}{\bibfnamefont{O.}~\bibnamefont{Gutfleisch}},
  \bibinfo{author}{\bibfnamefont{O.}~\bibnamefont{Eriksson}}, \bibnamefont{and}
  \bibinfo{author}{\bibfnamefont{L.}~\bibnamefont{Vitos}},
  \bibinfo{journal}{Sci. Rep.} \textbf{\bibinfo{volume}{11}},
  \bibinfo{pages}{1} (\bibinfo{year}{2021}).

\bibitem[{\citenamefont{Tuvshin et~al.}(2021)\citenamefont{Tuvshin,
  Ochirkhuyag, Hong, and Odkhuu}}]{Tuvshin-2021}
\bibinfo{author}{\bibfnamefont{D.}~\bibnamefont{Tuvshin}},
  \bibinfo{author}{\bibfnamefont{T.}~\bibnamefont{Ochirkhuyag}},
  \bibinfo{author}{\bibfnamefont{S.}~\bibnamefont{Hong}}, \bibnamefont{and}
  \bibinfo{author}{\bibfnamefont{D.}~\bibnamefont{Odkhuu}},
  \bibinfo{journal}{AIP Advances} \textbf{\bibinfo{volume}{11}},
  \bibinfo{pages}{015138} (\bibinfo{year}{2021}).

\bibitem[{\citenamefont{Paulev{\'e} et~al.}(1962)\citenamefont{Paulev{\'e},
  Dautreppe, Laugier, and N{\'e}el}}]{Pauleve-1962}
\bibinfo{author}{\bibfnamefont{J.}~\bibnamefont{Paulev{\'e}}},
  \bibinfo{author}{\bibfnamefont{D.}~\bibnamefont{Dautreppe}},
  \bibinfo{author}{\bibfnamefont{J.}~\bibnamefont{Laugier}}, \bibnamefont{and}
  \bibinfo{author}{\bibfnamefont{L.}~\bibnamefont{N{\'e}el}},
  \bibinfo{journal}{Compt. rend.} \textbf{\bibinfo{volume}{254}}
  (\bibinfo{year}{1962}).

\bibitem[{\citenamefont{Paulev{\'e} et~al.}(1968)\citenamefont{Paulev{\'e},
  Chamberod, Krebs, and Bourret}}]{Pauleve-1968}
\bibinfo{author}{\bibfnamefont{J.}~\bibnamefont{Paulev{\'e}}},
  \bibinfo{author}{\bibfnamefont{A.}~\bibnamefont{Chamberod}},
  \bibinfo{author}{\bibfnamefont{K.}~\bibnamefont{Krebs}}, \bibnamefont{and}
  \bibinfo{author}{\bibfnamefont{A.}~\bibnamefont{Bourret}},
  \bibinfo{journal}{J. Appl. Phys.} \textbf{\bibinfo{volume}{39}},
  \bibinfo{pages}{989} (\bibinfo{year}{1968}).

\bibitem[{\citenamefont{Poirier et~al.}(2015)\citenamefont{Poirier, Pinkerton,
  Kubic, Mishra, Bordeaux, Mubarok, Lewis, Goldstein, Skomski, and
  Barmak}}]{Poirier-2015}
\bibinfo{author}{\bibfnamefont{E.}~\bibnamefont{Poirier}},
  \bibinfo{author}{\bibfnamefont{F.~E.} \bibnamefont{Pinkerton}},
  \bibinfo{author}{\bibfnamefont{R.}~\bibnamefont{Kubic}},
  \bibinfo{author}{\bibfnamefont{R.~K.} \bibnamefont{Mishra}},
  \bibinfo{author}{\bibfnamefont{N.}~\bibnamefont{Bordeaux}},
  \bibinfo{author}{\bibfnamefont{A.}~\bibnamefont{Mubarok}},
  \bibinfo{author}{\bibfnamefont{L.~H.} \bibnamefont{Lewis}},
  \bibinfo{author}{\bibfnamefont{J.~I.} \bibnamefont{Goldstein}},
  \bibinfo{author}{\bibfnamefont{R.}~\bibnamefont{Skomski}}, \bibnamefont{and}
  \bibinfo{author}{\bibfnamefont{K.}~\bibnamefont{Barmak}},
  \bibinfo{journal}{J. Appl. Phys.} \textbf{\bibinfo{volume}{117}},
  \bibinfo{pages}{17E318} (\bibinfo{year}{2015}).

\bibitem[{\citenamefont{Skomski and Coey}(2019)}]{Skomski-2019permanent}
\bibinfo{author}{\bibfnamefont{R.}~\bibnamefont{Skomski}} \bibnamefont{and}
  \bibinfo{author}{\bibfnamefont{J.~M.~D.} \bibnamefont{Coey}},
  \emph{\bibinfo{title}{Permanent magnetism}} (\bibinfo{publisher}{Routledge},
  \bibinfo{year}{2019}).

\bibitem[{\citenamefont{Pastushenkov et~al.}(2005)\citenamefont{Pastushenkov,
  Skokov, Suponev, and Stakhovski}}]{Pastushenkov-2005}
\bibinfo{author}{\bibfnamefont{Y.~G.} \bibnamefont{Pastushenkov}},
  \bibinfo{author}{\bibfnamefont{K.}~\bibnamefont{Skokov}},
  \bibinfo{author}{\bibfnamefont{N.}~\bibnamefont{Suponev}}, \bibnamefont{and}
  \bibinfo{author}{\bibfnamefont{D.}~\bibnamefont{Stakhovski}},
  \bibinfo{journal}{J. Magn. Magn. Mater.} \textbf{\bibinfo{volume}{290}},
  \bibinfo{pages}{644} (\bibinfo{year}{2005}).

\bibitem[{\citenamefont{Otani et~al.}(1987)\citenamefont{Otani, Miyajima, and
  Chikazumi}}]{Otani-1987}
\bibinfo{author}{\bibfnamefont{Y.}~\bibnamefont{Otani}},
  \bibinfo{author}{\bibfnamefont{H.}~\bibnamefont{Miyajima}}, \bibnamefont{and}
  \bibinfo{author}{\bibfnamefont{S.}~\bibnamefont{Chikazumi}},
  \bibinfo{journal}{J. Appl. Phys.} \textbf{\bibinfo{volume}{61}},
  \bibinfo{pages}{3436} (\bibinfo{year}{1987}).

\bibitem[{\citenamefont{Lewis et~al.}(2014)\citenamefont{Lewis, Mubarok,
  Poirier, Bordeaux, Manchanda, Kashyap, Skomski, Goldstein, Pinkerton, Mishra
  et~al.}}]{Lewis-2014inspired}
\bibinfo{author}{\bibfnamefont{L.~H.} \bibnamefont{Lewis}},
  \bibinfo{author}{\bibfnamefont{A.}~\bibnamefont{Mubarok}},
  \bibinfo{author}{\bibfnamefont{E.}~\bibnamefont{Poirier}},
  \bibinfo{author}{\bibfnamefont{N.}~\bibnamefont{Bordeaux}},
  \bibinfo{author}{\bibfnamefont{P.}~\bibnamefont{Manchanda}},
  \bibinfo{author}{\bibfnamefont{A.}~\bibnamefont{Kashyap}},
  \bibinfo{author}{\bibfnamefont{R.}~\bibnamefont{Skomski}},
  \bibinfo{author}{\bibfnamefont{J.}~\bibnamefont{Goldstein}},
  \bibinfo{author}{\bibfnamefont{F.}~\bibnamefont{Pinkerton}},
  \bibinfo{author}{\bibfnamefont{R.}~\bibnamefont{Mishra}},
  \bibnamefont{et~al.}, \bibinfo{journal}{J. Phys. Condens. Mat.}
  \textbf{\bibinfo{volume}{26}}, \bibinfo{pages}{064213}
  (\bibinfo{year}{2014}).

\bibitem[{\citenamefont{N{\'e}el and Paulev{\'e}}(1964)}]{Neel-1964}
\bibinfo{author}{\bibfnamefont{L.}~\bibnamefont{N{\'e}el}} \bibnamefont{and}
  \bibinfo{author}{\bibfnamefont{J.}~\bibnamefont{Paulev{\'e}}},
  \bibinfo{journal}{J. Appl. Phys.} \textbf{\bibinfo{volume}{35}},
  \bibinfo{pages}{873} (\bibinfo{year}{1964}).

\bibitem[{\citenamefont{Lima~Jr and Drago}(2001)}]{Lima-2001}
\bibinfo{author}{\bibfnamefont{E.}~\bibnamefont{Lima~Jr}} \bibnamefont{and}
  \bibinfo{author}{\bibfnamefont{V.}~\bibnamefont{Drago}},
  \bibinfo{journal}{Phys. Stat. Sol. (A)} \textbf{\bibinfo{volume}{187}},
  \bibinfo{pages}{119} (\bibinfo{year}{2001}).

\bibitem[{\citenamefont{Shima et~al.}(2007)\citenamefont{Shima, Okamura,
  Mitani, and Takanashi}}]{Shima-2007}
\bibinfo{author}{\bibfnamefont{T.}~\bibnamefont{Shima}},
  \bibinfo{author}{\bibfnamefont{M.}~\bibnamefont{Okamura}},
  \bibinfo{author}{\bibfnamefont{S.}~\bibnamefont{Mitani}}, \bibnamefont{and}
  \bibinfo{author}{\bibfnamefont{K.}~\bibnamefont{Takanashi}},
  \bibinfo{journal}{J. Magn. Magn. Mater.} \textbf{\bibinfo{volume}{310}},
  \bibinfo{pages}{2213} (\bibinfo{year}{2007}).

\bibitem[{\citenamefont{Makino et~al.}(2015)\citenamefont{Makino, Sharma, Sato,
  Takeuchi, Zhang, and Takenaka}}]{Makino-2015}
\bibinfo{author}{\bibfnamefont{A.}~\bibnamefont{Makino}},
  \bibinfo{author}{\bibfnamefont{P.}~\bibnamefont{Sharma}},
  \bibinfo{author}{\bibfnamefont{K.}~\bibnamefont{Sato}},
  \bibinfo{author}{\bibfnamefont{A.}~\bibnamefont{Takeuchi}},
  \bibinfo{author}{\bibfnamefont{Y.}~\bibnamefont{Zhang}}, \bibnamefont{and}
  \bibinfo{author}{\bibfnamefont{K.}~\bibnamefont{Takenaka}},
  \bibinfo{journal}{Sci. Rep.} \textbf{\bibinfo{volume}{5}}, \bibinfo{pages}{1}
  (\bibinfo{year}{2015}).

\bibitem[{\citenamefont{Giannopoulos et~al.}(2018)\citenamefont{Giannopoulos,
  Barucca, Kaidatzis, Psycharis, Salikhov, Farle, Koutsouflakis, Niarchos,
  Mehta, Scuderi et~al.}}]{Giannopoulos-2018}
\bibinfo{author}{\bibfnamefont{G.}~\bibnamefont{Giannopoulos}},
  \bibinfo{author}{\bibfnamefont{G.}~\bibnamefont{Barucca}},
  \bibinfo{author}{\bibfnamefont{A.}~\bibnamefont{Kaidatzis}},
  \bibinfo{author}{\bibfnamefont{V.}~\bibnamefont{Psycharis}},
  \bibinfo{author}{\bibfnamefont{R.}~\bibnamefont{Salikhov}},
  \bibinfo{author}{\bibfnamefont{M.}~\bibnamefont{Farle}},
  \bibinfo{author}{\bibfnamefont{E.}~\bibnamefont{Koutsouflakis}},
  \bibinfo{author}{\bibfnamefont{D.}~\bibnamefont{Niarchos}},
  \bibinfo{author}{\bibfnamefont{A.}~\bibnamefont{Mehta}},
  \bibinfo{author}{\bibfnamefont{M.}~\bibnamefont{Scuderi}},
  \bibnamefont{et~al.}, \bibinfo{journal}{Sci. Rep.}
  \textbf{\bibinfo{volume}{8}}, \bibinfo{pages}{1} (\bibinfo{year}{2018}).

\bibitem[{\citenamefont{Goto et~al.}(2017)\citenamefont{Goto, Kura, Watanabe,
  Hayashi, Yanagihara, Shimada, Mizuguchi, Takanashi, and Kita}}]{Goto-2017}
\bibinfo{author}{\bibfnamefont{S.}~\bibnamefont{Goto}},
  \bibinfo{author}{\bibfnamefont{H.}~\bibnamefont{Kura}},
  \bibinfo{author}{\bibfnamefont{E.}~\bibnamefont{Watanabe}},
  \bibinfo{author}{\bibfnamefont{Y.}~\bibnamefont{Hayashi}},
  \bibinfo{author}{\bibfnamefont{H.}~\bibnamefont{Yanagihara}},
  \bibinfo{author}{\bibfnamefont{Y.}~\bibnamefont{Shimada}},
  \bibinfo{author}{\bibfnamefont{M.}~\bibnamefont{Mizuguchi}},
  \bibinfo{author}{\bibfnamefont{K.}~\bibnamefont{Takanashi}},
  \bibnamefont{and} \bibinfo{author}{\bibfnamefont{E.}~\bibnamefont{Kita}},
  \bibinfo{journal}{Sci. Rep.} \textbf{\bibinfo{volume}{7}}, \bibinfo{pages}{1}
  (\bibinfo{year}{2017}).

\bibitem[{\citenamefont{Izardar and Ederer}(2020)}]{Izardar-2020}
\bibinfo{author}{\bibfnamefont{A.}~\bibnamefont{Izardar}} \bibnamefont{and}
  \bibinfo{author}{\bibfnamefont{C.}~\bibnamefont{Ederer}},
  \bibinfo{journal}{Phys. Rev. Materials} \textbf{\bibinfo{volume}{4}},
  \bibinfo{pages}{054418} (\bibinfo{year}{2020}).

\bibitem[{\citenamefont{Matsushita et~al.}(2017)\citenamefont{Matsushita,
  Madjarova, Dewhurst, Shallcross, Felser, Sharma, and
  Gross}}]{Matsushita-2017}
\bibinfo{author}{\bibfnamefont{Y.}~\bibnamefont{Matsushita}},
  \bibinfo{author}{\bibfnamefont{G.}~\bibnamefont{Madjarova}},
  \bibinfo{author}{\bibfnamefont{J.}~\bibnamefont{Dewhurst}},
  \bibinfo{author}{\bibfnamefont{S.}~\bibnamefont{Shallcross}},
  \bibinfo{author}{\bibfnamefont{C.}~\bibnamefont{Felser}},
  \bibinfo{author}{\bibfnamefont{S.}~\bibnamefont{Sharma}}, \bibnamefont{and}
  \bibinfo{author}{\bibfnamefont{E.}~\bibnamefont{Gross}}, \bibinfo{journal}{J.
  Phys. D: Appl. Phys.} \textbf{\bibinfo{volume}{50}}, \bibinfo{pages}{095002}
  (\bibinfo{year}{2017}).

\bibitem[{\citenamefont{Herper}(2018)}]{Herper-2018}
\bibinfo{author}{\bibfnamefont{H.~C.} \bibnamefont{Herper}},
  \bibinfo{journal}{Phys. Rev. B} \textbf{\bibinfo{volume}{98}},
  \bibinfo{pages}{014411} (\bibinfo{year}{2018}).

\bibitem[{\citenamefont{Gao et~al.}(2020)\citenamefont{Gao, Opahle, Gutfleisch,
  and Zhang}}]{Gao-2020}
\bibinfo{author}{\bibfnamefont{Q.}~\bibnamefont{Gao}},
  \bibinfo{author}{\bibfnamefont{I.}~\bibnamefont{Opahle}},
  \bibinfo{author}{\bibfnamefont{O.}~\bibnamefont{Gutfleisch}},
  \bibnamefont{and} \bibinfo{author}{\bibfnamefont{H.}~\bibnamefont{Zhang}},
  \bibinfo{journal}{Acta Mater.} \textbf{\bibinfo{volume}{186}},
  \bibinfo{pages}{355} (\bibinfo{year}{2020}).

\bibitem[{\citenamefont{Zhang et~al.}(2013)\citenamefont{Zhang, Wang, Zhang,
  Liu, Ma, and Wu}}]{Zhang-2013}
\bibinfo{author}{\bibfnamefont{Y.}~\bibnamefont{Zhang}},
  \bibinfo{author}{\bibfnamefont{W.}~\bibnamefont{Wang}},
  \bibinfo{author}{\bibfnamefont{H.}~\bibnamefont{Zhang}},
  \bibinfo{author}{\bibfnamefont{E.}~\bibnamefont{Liu}},
  \bibinfo{author}{\bibfnamefont{R.}~\bibnamefont{Ma}}, \bibnamefont{and}
  \bibinfo{author}{\bibfnamefont{G.}~\bibnamefont{Wu}},
  \bibinfo{journal}{Physica B: Condens. Matter} \textbf{\bibinfo{volume}{420}},
  \bibinfo{pages}{86} (\bibinfo{year}{2013}).

\bibitem[{\citenamefont{Gupta and Bhat}(2014)}]{Gupta-2014}
\bibinfo{author}{\bibfnamefont{D.~C.} \bibnamefont{Gupta}} \bibnamefont{and}
  \bibinfo{author}{\bibfnamefont{I.~H.} \bibnamefont{Bhat}},
  \bibinfo{journal}{Mater. Chem. Phys.} \textbf{\bibinfo{volume}{146}},
  \bibinfo{pages}{303} (\bibinfo{year}{2014}).

\bibitem[{\citenamefont{Dahmane et~al.}(2016)\citenamefont{Dahmane, Mogulkoc,
  Doumi, Tadjer, Khenata, Omran, Rai, Murtaza, and Varshney}}]{Dahmane-2016}
\bibinfo{author}{\bibfnamefont{F.}~\bibnamefont{Dahmane}},
  \bibinfo{author}{\bibfnamefont{Y.}~\bibnamefont{Mogulkoc}},
  \bibinfo{author}{\bibfnamefont{B.}~\bibnamefont{Doumi}},
  \bibinfo{author}{\bibfnamefont{A.}~\bibnamefont{Tadjer}},
  \bibinfo{author}{\bibfnamefont{R.}~\bibnamefont{Khenata}},
  \bibinfo{author}{\bibfnamefont{S.~B.} \bibnamefont{Omran}},
  \bibinfo{author}{\bibfnamefont{D.}~\bibnamefont{Rai}},
  \bibinfo{author}{\bibfnamefont{G.}~\bibnamefont{Murtaza}}, \bibnamefont{and}
  \bibinfo{author}{\bibfnamefont{D.}~\bibnamefont{Varshney}},
  \bibinfo{journal}{J. Magn. Magn. Mater.} \textbf{\bibinfo{volume}{407}},
  \bibinfo{pages}{167} (\bibinfo{year}{2016}).

\bibitem[{\citenamefont{Popiel et~al.}(2004)\citenamefont{Popiel, Zarek, and
  Tuszy{\'n}ski}}]{Popiel-2004}
\bibinfo{author}{\bibfnamefont{E.~S.} \bibnamefont{Popiel}},
  \bibinfo{author}{\bibfnamefont{W.}~\bibnamefont{Zarek}}, \bibnamefont{and}
  \bibinfo{author}{\bibfnamefont{M.}~\bibnamefont{Tuszy{\'n}ski}},
  \bibinfo{journal}{Nukleonika} \textbf{\bibinfo{volume}{49}},
  \bibinfo{pages}{49} (\bibinfo{year}{2004}).

\bibitem[{\citenamefont{Yin et~al.}(2015)\citenamefont{Yin, Nash, and
  Chen}}]{Yin-2015}
\bibinfo{author}{\bibfnamefont{M.}~\bibnamefont{Yin}},
  \bibinfo{author}{\bibfnamefont{P.}~\bibnamefont{Nash}}, \bibnamefont{and}
  \bibinfo{author}{\bibfnamefont{S.}~\bibnamefont{Chen}},
  \bibinfo{journal}{Intermetallics} \textbf{\bibinfo{volume}{57}},
  \bibinfo{pages}{34} (\bibinfo{year}{2015}).

\bibitem[{\citenamefont{Zhang et~al.}(2009)\citenamefont{Zhang, Wang, Du, Hu,
  Nash, Lu, and Jiang}}]{Zhang-2009}
\bibinfo{author}{\bibfnamefont{L.}~\bibnamefont{Zhang}},
  \bibinfo{author}{\bibfnamefont{J.}~\bibnamefont{Wang}},
  \bibinfo{author}{\bibfnamefont{Y.}~\bibnamefont{Du}},
  \bibinfo{author}{\bibfnamefont{R.}~\bibnamefont{Hu}},
  \bibinfo{author}{\bibfnamefont{P.}~\bibnamefont{Nash}},
  \bibinfo{author}{\bibfnamefont{X.-G.} \bibnamefont{Lu}}, \bibnamefont{and}
  \bibinfo{author}{\bibfnamefont{C.}~\bibnamefont{Jiang}},
  \bibinfo{journal}{Acta Mater.} \textbf{\bibinfo{volume}{57}},
  \bibinfo{pages}{5324} (\bibinfo{year}{2009}).

\bibitem[{\citenamefont{Bradley and Taylor}(1937)}]{Bradley-1937}
\bibinfo{author}{\bibfnamefont{A.}~\bibnamefont{Bradley}} \bibnamefont{and}
  \bibinfo{author}{\bibfnamefont{A.}~\bibnamefont{Taylor}},
  \bibinfo{journal}{Nature} \textbf{\bibinfo{volume}{140}},
  \bibinfo{pages}{1012} (\bibinfo{year}{1937}).

\bibitem[{\citenamefont{Menushenkov et~al.}(2017)\citenamefont{Menushenkov,
  Gorshenkov, Savchenko, Shchetinin, and Savchenko}}]{Menushenkov-2017}
\bibinfo{author}{\bibfnamefont{V.}~\bibnamefont{Menushenkov}},
  \bibinfo{author}{\bibfnamefont{M.}~\bibnamefont{Gorshenkov}},
  \bibinfo{author}{\bibfnamefont{E.}~\bibnamefont{Savchenko}},
  \bibinfo{author}{\bibfnamefont{I.}~\bibnamefont{Shchetinin}},
  \bibnamefont{and}
  \bibinfo{author}{\bibfnamefont{A.}~\bibnamefont{Savchenko}},
  \bibinfo{journal}{Heat Treat. Met.} \textbf{\bibinfo{volume}{59}},
  \bibinfo{pages}{518} (\bibinfo{year}{2017}).

\bibitem[{\citenamefont{Hao et~al.}(1984)\citenamefont{Hao, Takayama, Ishida,
  and Nishizawa}}]{Hao-1984}
\bibinfo{author}{\bibfnamefont{S.~M.} \bibnamefont{Hao}},
  \bibinfo{author}{\bibfnamefont{T.}~\bibnamefont{Takayama}},
  \bibinfo{author}{\bibfnamefont{K.}~\bibnamefont{Ishida}}, \bibnamefont{and}
  \bibinfo{author}{\bibfnamefont{T.}~\bibnamefont{Nishizawa}},
  \bibinfo{journal}{Metall. Trans. A} \textbf{\bibinfo{volume}{15}},
  \bibinfo{pages}{1819} (\bibinfo{year}{1984}).

\bibitem[{\citenamefont{Zhang et~al.}(2008)\citenamefont{Zhang, Du, Xu, Tang,
  Chen, and Zhang}}]{Zhang-2008}
\bibinfo{author}{\bibfnamefont{L.}~\bibnamefont{Zhang}},
  \bibinfo{author}{\bibfnamefont{Y.}~\bibnamefont{Du}},
  \bibinfo{author}{\bibfnamefont{H.}~\bibnamefont{Xu}},
  \bibinfo{author}{\bibfnamefont{C.}~\bibnamefont{Tang}},
  \bibinfo{author}{\bibfnamefont{H.}~\bibnamefont{Chen}}, \bibnamefont{and}
  \bibinfo{author}{\bibfnamefont{W.}~\bibnamefont{Zhang}}, \bibinfo{journal}{J.
  Alloys Compd.} \textbf{\bibinfo{volume}{454}}, \bibinfo{pages}{129}
  (\bibinfo{year}{2008}).

\bibitem[{\citenamefont{Buschow and de~Boer}(2007)}]{Buschow-2007}
\bibinfo{author}{\bibfnamefont{K.}~\bibnamefont{Buschow}} \bibnamefont{and}
  \bibinfo{author}{\bibfnamefont{F.}~\bibnamefont{de~Boer}},
  \emph{\bibinfo{title}{Physics of Magnetism and Magnetic Materials}}, Focus on
  biotechnology (\bibinfo{publisher}{Springer US}, \bibinfo{year}{2007}).

\bibitem[{\citenamefont{Menushenkov
  et~al.}(2015{\natexlab{a}})\citenamefont{Menushenkov, Gorshenkov, Zhukov,
  Savchenko, and Zheleznyi}}]{Menushenkov-2015a}
\bibinfo{author}{\bibfnamefont{V.}~\bibnamefont{Menushenkov}},
  \bibinfo{author}{\bibfnamefont{M.}~\bibnamefont{Gorshenkov}},
  \bibinfo{author}{\bibfnamefont{D.}~\bibnamefont{Zhukov}},
  \bibinfo{author}{\bibfnamefont{E.}~\bibnamefont{Savchenko}},
  \bibnamefont{and}
  \bibinfo{author}{\bibfnamefont{M.}~\bibnamefont{Zheleznyi}},
  \bibinfo{journal}{Mater. Lett.} \textbf{\bibinfo{volume}{152}},
  \bibinfo{pages}{68} (\bibinfo{year}{2015}{\natexlab{a}}).

\bibitem[{\citenamefont{Menushenkov
  et~al.}(2015{\natexlab{b}})\citenamefont{Menushenkov, Gorshenkov, Shchetinin,
  Savchenko, Savchenko, and Zhukov}}]{Menushenkov-2015b}
\bibinfo{author}{\bibfnamefont{V.}~\bibnamefont{Menushenkov}},
  \bibinfo{author}{\bibfnamefont{M.}~\bibnamefont{Gorshenkov}},
  \bibinfo{author}{\bibfnamefont{I.}~\bibnamefont{Shchetinin}},
  \bibinfo{author}{\bibfnamefont{A.}~\bibnamefont{Savchenko}},
  \bibinfo{author}{\bibfnamefont{E.}~\bibnamefont{Savchenko}},
  \bibnamefont{and} \bibinfo{author}{\bibfnamefont{D.}~\bibnamefont{Zhukov}},
  \bibinfo{journal}{J. Magn. Magn. Mater.} \textbf{\bibinfo{volume}{390}},
  \bibinfo{pages}{40} (\bibinfo{year}{2015}{\natexlab{b}}).

\bibitem[{\citenamefont{Kresse and Furthm\"uller}(1996)}]{Kresse-1996}
\bibinfo{author}{\bibfnamefont{G.}~\bibnamefont{Kresse}} \bibnamefont{and}
  \bibinfo{author}{\bibfnamefont{J.}~\bibnamefont{Furthm\"uller}},
  \bibinfo{journal}{Phys. Rev. B} \textbf{\bibinfo{volume}{54}},
  \bibinfo{pages}{11169} (\bibinfo{year}{1996}).

\bibitem[{\citenamefont{Perdew et~al.}(1996)\citenamefont{Perdew, Burke, and
  Ernzerhof}}]{Perdew1996}
\bibinfo{author}{\bibfnamefont{J.~P.} \bibnamefont{Perdew}},
  \bibinfo{author}{\bibfnamefont{K.}~\bibnamefont{Burke}}, \bibnamefont{and}
  \bibinfo{author}{\bibfnamefont{M.}~\bibnamefont{Ernzerhof}},
  \bibinfo{journal}{Phys. Rev. Lett.} \textbf{\bibinfo{volume}{77}},
  \bibinfo{pages}{3865} (\bibinfo{year}{1996}).

\bibitem[{\citenamefont{Neibecker et~al.}(2017)\citenamefont{Neibecker, Gruner,
  Xu, Kainuma, Petry, Pentcheva, and Leitner}}]{Neibecker-2017}
\bibinfo{author}{\bibfnamefont{P.}~\bibnamefont{Neibecker}},
  \bibinfo{author}{\bibfnamefont{M.~E.} \bibnamefont{Gruner}},
  \bibinfo{author}{\bibfnamefont{X.}~\bibnamefont{Xu}},
  \bibinfo{author}{\bibfnamefont{R.}~\bibnamefont{Kainuma}},
  \bibinfo{author}{\bibfnamefont{W.}~\bibnamefont{Petry}},
  \bibinfo{author}{\bibfnamefont{R.}~\bibnamefont{Pentcheva}},
  \bibnamefont{and} \bibinfo{author}{\bibfnamefont{M.}~\bibnamefont{Leitner}},
  \bibinfo{journal}{Phys. Rev. B} \textbf{\bibinfo{volume}{96}},
  \bibinfo{pages}{165131} (\bibinfo{year}{2017}).

\bibitem[{SM()}]{SM}
\bibinfo{note}{See Supplemental Material at http://... for additional
  information about MAE convergence test, crystal structures for
  off-stoichiometric compositions, decomposition energies, and SOC contribution
  to the~MAE.}

\bibitem[{\citenamefont{Liechtenstein et~al.}(1984)\citenamefont{Liechtenstein,
  Katsnelson, and Gubanov}}]{Liechtenstein-1984}
\bibinfo{author}{\bibfnamefont{A.}~\bibnamefont{Liechtenstein}},
  \bibinfo{author}{\bibfnamefont{M.}~\bibnamefont{Katsnelson}},
  \bibnamefont{and} \bibinfo{author}{\bibfnamefont{V.}~\bibnamefont{Gubanov}},
  \bibinfo{journal}{J. Phys. F: Met. Phys.} \textbf{\bibinfo{volume}{14}},
  \bibinfo{pages}{L125} (\bibinfo{year}{1984}).

\bibitem[{\citenamefont{Liechtenstein et~al.}(1985)\citenamefont{Liechtenstein,
  Katsnelson, and Gubanov}}]{Liechtenstein-1985}
\bibinfo{author}{\bibfnamefont{A.}~\bibnamefont{Liechtenstein}},
  \bibinfo{author}{\bibfnamefont{M.}~\bibnamefont{Katsnelson}},
  \bibnamefont{and} \bibinfo{author}{\bibfnamefont{V.}~\bibnamefont{Gubanov}},
  \bibinfo{journal}{Solid State Commun.} \textbf{\bibinfo{volume}{54}},
  \bibinfo{pages}{327} (\bibinfo{year}{1985}).

\bibitem[{\citenamefont{Liechtenstein et~al.}(1987)\citenamefont{Liechtenstein,
  Katsnelson, Antropov, and Gubanov}}]{Liechtenstein-1987}
\bibinfo{author}{\bibfnamefont{A.~I.} \bibnamefont{Liechtenstein}},
  \bibinfo{author}{\bibfnamefont{M.}~\bibnamefont{Katsnelson}},
  \bibinfo{author}{\bibfnamefont{V.}~\bibnamefont{Antropov}}, \bibnamefont{and}
  \bibinfo{author}{\bibfnamefont{V.}~\bibnamefont{Gubanov}},
  \bibinfo{journal}{J. Magn. Magn. Mater.} \textbf{\bibinfo{volume}{67}},
  \bibinfo{pages}{65} (\bibinfo{year}{1987}).

\bibitem[{\citenamefont{Enkovaara et~al.}(2002)\citenamefont{Enkovaara, Ayuela,
  Nordstr{\"o}m, and Nieminen}}]{Enkovaara-2002}
\bibinfo{author}{\bibfnamefont{J.}~\bibnamefont{Enkovaara}},
  \bibinfo{author}{\bibfnamefont{A.}~\bibnamefont{Ayuela}},
  \bibinfo{author}{\bibfnamefont{L.}~\bibnamefont{Nordstr{\"o}m}},
  \bibnamefont{and} \bibinfo{author}{\bibfnamefont{R.~M.}
  \bibnamefont{Nieminen}}, \bibinfo{journal}{Phys. Rev. B}
  \textbf{\bibinfo{volume}{65}}, \bibinfo{pages}{134422}
  (\bibinfo{year}{2002}).

\bibitem[{\citenamefont{Umetsu et~al.}(2006)\citenamefont{Umetsu, Sakuma, and
  Fukamichi}}]{Umetsu-2006}
\bibinfo{author}{\bibfnamefont{R.}~\bibnamefont{Umetsu}},
  \bibinfo{author}{\bibfnamefont{A.}~\bibnamefont{Sakuma}}, \bibnamefont{and}
  \bibinfo{author}{\bibfnamefont{K.}~\bibnamefont{Fukamichi}},
  \bibinfo{journal}{Appl. Phys. Lett.} \textbf{\bibinfo{volume}{89}},
  \bibinfo{pages}{052504} (\bibinfo{year}{2006}).

\bibitem[{\citenamefont{Gruner et~al.}(2008)\citenamefont{Gruner, Entel,
  Opahle, and Richter}}]{Gruner-2008}
\bibinfo{author}{\bibfnamefont{M.~E.} \bibnamefont{Gruner}},
  \bibinfo{author}{\bibfnamefont{P.}~\bibnamefont{Entel}},
  \bibinfo{author}{\bibfnamefont{I.}~\bibnamefont{Opahle}}, \bibnamefont{and}
  \bibinfo{author}{\bibfnamefont{M.}~\bibnamefont{Richter}},
  \bibinfo{journal}{J. Mater. Sci.} \textbf{\bibinfo{volume}{43}},
  \bibinfo{pages}{3825} (\bibinfo{year}{2008}).

\bibitem[{\citenamefont{Edstr{\"o}m et~al.}(2015)\citenamefont{Edstr{\"o}m,
  Werwi{\'n}ski, Iu{\c{s}}an, Rusz, Eriksson, Skokov, Radulov, Ener, Kuz'Min,
  Hong et~al.}}]{Edstrom-2015}
\bibinfo{author}{\bibfnamefont{A.}~\bibnamefont{Edstr{\"o}m}},
  \bibinfo{author}{\bibfnamefont{M.}~\bibnamefont{Werwi{\'n}ski}},
  \bibinfo{author}{\bibfnamefont{D.}~\bibnamefont{Iu{\c{s}}an}},
  \bibinfo{author}{\bibfnamefont{J.}~\bibnamefont{Rusz}},
  \bibinfo{author}{\bibfnamefont{O.}~\bibnamefont{Eriksson}},
  \bibinfo{author}{\bibfnamefont{K.}~\bibnamefont{Skokov}},
  \bibinfo{author}{\bibfnamefont{I.}~\bibnamefont{Radulov}},
  \bibinfo{author}{\bibfnamefont{S.}~\bibnamefont{Ener}},
  \bibinfo{author}{\bibfnamefont{M.}~\bibnamefont{Kuz'Min}},
  \bibinfo{author}{\bibfnamefont{J.}~\bibnamefont{Hong}}, \bibnamefont{et~al.},
  \bibinfo{journal}{Phys. Rev. B} \textbf{\bibinfo{volume}{92}},
  \bibinfo{pages}{174413} (\bibinfo{year}{2015}).

\bibitem[{\citenamefont{Togo and Tanaka}(2015)}]{Phonopy}
\bibinfo{author}{\bibfnamefont{A.}~\bibnamefont{Togo}} \bibnamefont{and}
  \bibinfo{author}{\bibfnamefont{I.}~\bibnamefont{Tanaka}},
  \bibinfo{journal}{Scr. Mater.} \textbf{\bibinfo{volume}{108}},
  \bibinfo{pages}{1} (\bibinfo{year}{2015}).

\bibitem[{\citenamefont{Ebert et~al.}(2011)\citenamefont{Ebert,
  K\"{o}dderitzsch, and Min\'{a}r}}]{Ebert-2011}
\bibinfo{author}{\bibfnamefont{H.}~\bibnamefont{Ebert}},
  \bibinfo{author}{\bibfnamefont{D.}~\bibnamefont{K\"{o}dderitzsch}},
  \bibnamefont{and}
  \bibinfo{author}{\bibfnamefont{J.}~\bibnamefont{Min\'{a}r}},
  \bibinfo{journal}{Rep. Prog. Phys.} \textbf{\bibinfo{volume}{74}},
  \bibinfo{pages}{096501} (\bibinfo{year}{2011}).

\bibitem[{x05()}]{x05-cell}
\bibinfo{note}{Due to the distribution of the atoms, a 16 atom cell of
  Fe$_2$Ni$_{1.5}$Al$_{0.5}$ (Fe$_8$Ni$_6$Al$_2$) possesses two inequivalent
  crystallographic directions along $x$ or $y$ and $z$ axes even with cubic
  lattice parameters $a=b=c$. Replacing two Al in the inverse Heusler
  supercell, we necessarily obtain one atomic layer which completely consists
  of Ni atoms. As~a result, a tetragonal distortion can be applied to the
  lattice parameters which are parallel or orthogonal to this layer.
  The~respective results are averaged.}

\bibitem[{\citenamefont{Siewert et~al.}(2011)\citenamefont{Siewert, Gruner,
  Dannenberg, Chakrabarti, Herper, Wuttig, Barman, Singh, Al-Zubi, Hickel
  et~al.}}]{Siewert-2011}
\bibinfo{author}{\bibfnamefont{M.}~\bibnamefont{Siewert}},
  \bibinfo{author}{\bibfnamefont{M.~E.} \bibnamefont{Gruner}},
  \bibinfo{author}{\bibfnamefont{A.}~\bibnamefont{Dannenberg}},
  \bibinfo{author}{\bibfnamefont{A.}~\bibnamefont{Chakrabarti}},
  \bibinfo{author}{\bibfnamefont{H.~C.} \bibnamefont{Herper}},
  \bibinfo{author}{\bibfnamefont{M.}~\bibnamefont{Wuttig}},
  \bibinfo{author}{\bibfnamefont{S.~R.} \bibnamefont{Barman}},
  \bibinfo{author}{\bibfnamefont{S.}~\bibnamefont{Singh}},
  \bibinfo{author}{\bibfnamefont{A.}~\bibnamefont{Al-Zubi}},
  \bibinfo{author}{\bibfnamefont{T.}~\bibnamefont{Hickel}},
  \bibnamefont{et~al.}, \bibinfo{journal}{Appl. Phys. Lett.}
  \textbf{\bibinfo{volume}{99}}, \bibinfo{pages}{191904}
  (\bibinfo{year}{2011}).

\bibitem[{\citenamefont{Wu and Freeman}(1999)}]{Wu-1999}
\bibinfo{author}{\bibfnamefont{R.}~\bibnamefont{Wu}} \bibnamefont{and}
  \bibinfo{author}{\bibfnamefont{A.}~\bibnamefont{Freeman}},
  \bibinfo{journal}{J. Magn. Magn. Mater.} \textbf{\bibinfo{volume}{200}},
  \bibinfo{pages}{498} (\bibinfo{year}{1999}).

\bibitem[{\citenamefont{Miura et~al.}(2013)\citenamefont{Miura, Ozaki,
  Kuwahara, Tsujikawa, Abe, and Shirai}}]{Miura-2013}
\bibinfo{author}{\bibfnamefont{Y.}~\bibnamefont{Miura}},
  \bibinfo{author}{\bibfnamefont{S.}~\bibnamefont{Ozaki}},
  \bibinfo{author}{\bibfnamefont{Y.}~\bibnamefont{Kuwahara}},
  \bibinfo{author}{\bibfnamefont{M.}~\bibnamefont{Tsujikawa}},
  \bibinfo{author}{\bibfnamefont{K.}~\bibnamefont{Abe}}, \bibnamefont{and}
  \bibinfo{author}{\bibfnamefont{M.}~\bibnamefont{Shirai}},
  \bibinfo{journal}{J. Phys. Condens. Matter.} \textbf{\bibinfo{volume}{25}},
  \bibinfo{pages}{106005} (\bibinfo{year}{2013}).

\bibitem[{\citenamefont{{Lewis} et~al.}(2014)\citenamefont{{Lewis},
  {Pinkerton}, {Bordeaux}, {Mubarok}, {Poirier}, {Goldstein}, {Skomski}, and
  {Barmak}}}]{Lewis-2014}
\bibinfo{author}{\bibfnamefont{L.~H.} \bibnamefont{{Lewis}}},
  \bibinfo{author}{\bibfnamefont{F.~E.} \bibnamefont{{Pinkerton}}},
  \bibinfo{author}{\bibfnamefont{N.}~\bibnamefont{{Bordeaux}}},
  \bibinfo{author}{\bibfnamefont{A.}~\bibnamefont{{Mubarok}}},
  \bibinfo{author}{\bibfnamefont{E.}~\bibnamefont{{Poirier}}},
  \bibinfo{author}{\bibfnamefont{J.~I.} \bibnamefont{{Goldstein}}},
  \bibinfo{author}{\bibfnamefont{R.}~\bibnamefont{{Skomski}}},
  \bibnamefont{and} \bibinfo{author}{\bibfnamefont{K.}~\bibnamefont{{Barmak}}},
  \bibinfo{journal}{IEEE Magn. Lett.} \textbf{\bibinfo{volume}{5}},
  \bibinfo{pages}{1} (\bibinfo{year}{2014}).

\bibitem[{\citenamefont{Bruno}(1989)}]{Bruno-1989}
\bibinfo{author}{\bibfnamefont{P.}~\bibnamefont{Bruno}},
  \bibinfo{journal}{Phys. Rev. B} \textbf{\bibinfo{volume}{39}},
  \bibinfo{pages}{865} (\bibinfo{year}{1989}).

\bibitem[{\citenamefont{Wilhelm et~al.}(2001)\citenamefont{Wilhelm,
  Poulopoulos, Wende, Scherz, Baberschke, Angelakeris, Flevaris, and
  Rogalev}}]{Wilhelm-2001}
\bibinfo{author}{\bibfnamefont{F.}~\bibnamefont{Wilhelm}},
  \bibinfo{author}{\bibfnamefont{P.}~\bibnamefont{Poulopoulos}},
  \bibinfo{author}{\bibfnamefont{H.}~\bibnamefont{Wende}},
  \bibinfo{author}{\bibfnamefont{A.}~\bibnamefont{Scherz}},
  \bibinfo{author}{\bibfnamefont{K.}~\bibnamefont{Baberschke}},
  \bibinfo{author}{\bibfnamefont{M.}~\bibnamefont{Angelakeris}},
  \bibinfo{author}{\bibfnamefont{N.~K.} \bibnamefont{Flevaris}},
  \bibnamefont{and} \bibinfo{author}{\bibfnamefont{A.}~\bibnamefont{Rogalev}},
  \bibinfo{journal}{Phys. Rev. Lett.} \textbf{\bibinfo{volume}{87}},
  \bibinfo{pages}{207202} (\bibinfo{year}{2001}).

\bibitem[{\citenamefont{Andersson et~al.}(2007)\citenamefont{Andersson, Sanyal,
  Eriksson, Nordstr\"om, Karis, Arvanitis, Konishi, Holub-Krappe, and
  Dunn}}]{Andersson-2007}
\bibinfo{author}{\bibfnamefont{C.}~\bibnamefont{Andersson}},
  \bibinfo{author}{\bibfnamefont{B.}~\bibnamefont{Sanyal}},
  \bibinfo{author}{\bibfnamefont{O.}~\bibnamefont{Eriksson}},
  \bibinfo{author}{\bibfnamefont{L.}~\bibnamefont{Nordstr\"om}},
  \bibinfo{author}{\bibfnamefont{O.}~\bibnamefont{Karis}},
  \bibinfo{author}{\bibfnamefont{D.}~\bibnamefont{Arvanitis}},
  \bibinfo{author}{\bibfnamefont{T.}~\bibnamefont{Konishi}},
  \bibinfo{author}{\bibfnamefont{E.}~\bibnamefont{Holub-Krappe}},
  \bibnamefont{and} \bibinfo{author}{\bibfnamefont{J.}~\bibnamefont{Dunn}},
  \bibinfo{journal}{Phys. Rev. Lett.} \textbf{\bibinfo{volume}{99}},
  \bibinfo{pages}{177207} (\bibinfo{year}{2007}).

\bibitem[{\citenamefont{Solovyev et~al.}(1995)\citenamefont{Solovyev,
  Dederichs, and Mertig}}]{Solovyev-1995}
\bibinfo{author}{\bibfnamefont{I.~V.} \bibnamefont{Solovyev}},
  \bibinfo{author}{\bibfnamefont{P.~H.} \bibnamefont{Dederichs}},
  \bibnamefont{and} \bibinfo{author}{\bibfnamefont{I.}~\bibnamefont{Mertig}},
  \bibinfo{journal}{Phys. Rev. B} \textbf{\bibinfo{volume}{52}},
  \bibinfo{pages}{13419} (\bibinfo{year}{1995}).

\bibitem[{\citenamefont{Ravindran et~al.}(2001)\citenamefont{Ravindran,
  Kjekshus, Fjellvaag, James, Nordstr\"om, Johansson, and
  Eriksson}}]{Ravindran-2001}
\bibinfo{author}{\bibfnamefont{P.}~\bibnamefont{Ravindran}},
  \bibinfo{author}{\bibfnamefont{A.}~\bibnamefont{Kjekshus}},
  \bibinfo{author}{\bibfnamefont{H.}~\bibnamefont{Fjellvaag}},
  \bibinfo{author}{\bibfnamefont{P.}~\bibnamefont{James}},
  \bibinfo{author}{\bibfnamefont{L.}~\bibnamefont{Nordstr\"om}},
  \bibinfo{author}{\bibfnamefont{B.}~\bibnamefont{Johansson}},
  \bibnamefont{and} \bibinfo{author}{\bibfnamefont{O.}~\bibnamefont{Eriksson}},
  \bibinfo{journal}{Phys. Rev. B} \textbf{\bibinfo{volume}{63}},
  \bibinfo{pages}{144409} (\bibinfo{year}{2001}).

\bibitem[{\citenamefont{Gruner}(2013)}]{Gruner-2013}
\bibinfo{author}{\bibfnamefont{M.~E.} \bibnamefont{Gruner}},
  \bibinfo{journal}{Phys. Stat. Sol. (A)} \textbf{\bibinfo{volume}{210}},
  \bibinfo{pages}{1282} (\bibinfo{year}{2013}).

\bibitem[{\citenamefont{Wolloch et~al.}(2017)\citenamefont{Wolloch, Suess, and
  Mohn}}]{Wolloch-2017}
\bibinfo{author}{\bibfnamefont{M.}~\bibnamefont{Wolloch}},
  \bibinfo{author}{\bibfnamefont{D.}~\bibnamefont{Suess}}, \bibnamefont{and}
  \bibinfo{author}{\bibfnamefont{P.}~\bibnamefont{Mohn}},
  \bibinfo{journal}{Phys. Rev. B} \textbf{\bibinfo{volume}{96}},
  \bibinfo{pages}{104408} (\bibinfo{year}{2017}).

\bibitem[{\citenamefont{Enkovaara et~al.}(2003)\citenamefont{Enkovaara, Ayuela,
  Jalkanen, Nordstr\"om, and Nieminen}}]{Enkovaara-2003}
\bibinfo{author}{\bibfnamefont{J.}~\bibnamefont{Enkovaara}},
  \bibinfo{author}{\bibfnamefont{A.}~\bibnamefont{Ayuela}},
  \bibinfo{author}{\bibfnamefont{J.}~\bibnamefont{Jalkanen}},
  \bibinfo{author}{\bibfnamefont{L.}~\bibnamefont{Nordstr\"om}},
  \bibnamefont{and} \bibinfo{author}{\bibfnamefont{R.~M.}
  \bibnamefont{Nieminen}}, \bibinfo{journal}{Phys. Rev. B}
  \textbf{\bibinfo{volume}{67}}, \bibinfo{pages}{054417}
  (\bibinfo{year}{2003}).

\bibitem[{\citenamefont{Garanin}(1996)}]{Garanin-1996}
\bibinfo{author}{\bibfnamefont{D.}~\bibnamefont{Garanin}},
  \bibinfo{journal}{Phys. Rev. B} \textbf{\bibinfo{volume}{53}},
  \bibinfo{pages}{11593} (\bibinfo{year}{1996}).

\bibitem[{\citenamefont{Sokolovskiy et~al.}(2012)\citenamefont{Sokolovskiy,
  Buchelnikov, Zagrebin, Entel, Sahoo, and Ogura}}]{Sokolovskiy-2012}
\bibinfo{author}{\bibfnamefont{V.}~\bibnamefont{Sokolovskiy}},
  \bibinfo{author}{\bibfnamefont{V.}~\bibnamefont{Buchelnikov}},
  \bibinfo{author}{\bibfnamefont{M.}~\bibnamefont{Zagrebin}},
  \bibinfo{author}{\bibfnamefont{P.}~\bibnamefont{Entel}},
  \bibinfo{author}{\bibfnamefont{S.}~\bibnamefont{Sahoo}}, \bibnamefont{and}
  \bibinfo{author}{\bibfnamefont{M.}~\bibnamefont{Ogura}},
  \bibinfo{journal}{Phys. Rev. B} \textbf{\bibinfo{volume}{86}},
  \bibinfo{pages}{134418} (\bibinfo{year}{2012}).

\bibitem[{\citenamefont{Meinert}(2016)}]{Meinert-2016}
\bibinfo{author}{\bibfnamefont{M.}~\bibnamefont{Meinert}}, \bibinfo{journal}{J.
  Phys. Condens. Matter} \textbf{\bibinfo{volume}{28}}, \bibinfo{pages}{056006}
  (\bibinfo{year}{2016}).

\bibitem[{\citenamefont{Wasilewski et~al.}(2018)\citenamefont{Wasilewski,
  Marciniak, and Werwi{\'n}ski}}]{Wasilewski-2018}
\bibinfo{author}{\bibfnamefont{B.}~\bibnamefont{Wasilewski}},
  \bibinfo{author}{\bibfnamefont{W.}~\bibnamefont{Marciniak}},
  \bibnamefont{and}
  \bibinfo{author}{\bibfnamefont{M.}~\bibnamefont{Werwi{\'n}ski}},
  \bibinfo{journal}{J. Phys. D: Appl. Phys.} \textbf{\bibinfo{volume}{51}},
  \bibinfo{pages}{175001} (\bibinfo{year}{2018}).

\bibitem[{\citenamefont{Wei and Zhou}(2018)}]{Wei-2018}
\bibinfo{author}{\bibfnamefont{X.-P.} \bibnamefont{Wei}} \bibnamefont{and}
  \bibinfo{author}{\bibfnamefont{Y.-H.} \bibnamefont{Zhou}},
  \bibinfo{journal}{Intermetallics} \textbf{\bibinfo{volume}{93}},
  \bibinfo{pages}{283} (\bibinfo{year}{2018}).

\bibitem[{\citenamefont{Zagrebin et~al.}(2020)\citenamefont{Zagrebin,
  Matyunina, Miroshkina, Pavlukhina, Sokolovskiy, and
  Buchelnikov}}]{Zagrebin-2020}
\bibinfo{author}{\bibfnamefont{M.}~\bibnamefont{Zagrebin}},
  \bibinfo{author}{\bibfnamefont{M.}~\bibnamefont{Matyunina}},
  \bibinfo{author}{\bibfnamefont{O.}~\bibnamefont{Miroshkina}},
  \bibinfo{author}{\bibfnamefont{O.}~\bibnamefont{Pavlukhina}},
  \bibinfo{author}{\bibfnamefont{V.}~\bibnamefont{Sokolovskiy}},
  \bibnamefont{and}
  \bibinfo{author}{\bibfnamefont{V.}~\bibnamefont{Buchelnikov}},
  \bibinfo{journal}{Ph. Transit.} \textbf{\bibinfo{volume}{93}},
  \bibinfo{pages}{43} (\bibinfo{year}{2020}).

\bibitem[{\citenamefont{Saito and Nishio-Hamane}(2018)}]{Saito-2018}
\bibinfo{author}{\bibfnamefont{T.}~\bibnamefont{Saito}} \bibnamefont{and}
  \bibinfo{author}{\bibfnamefont{D.}~\bibnamefont{Nishio-Hamane}},
  \bibinfo{journal}{J. Appl. Phys.} \textbf{\bibinfo{volume}{124}},
  \bibinfo{pages}{075105} (\bibinfo{year}{2018}).

\bibitem[{\citenamefont{Stanek et~al.}(2010)\citenamefont{Stanek, Wierzbicki,
  and Leonowicz}}]{Stanek-2010}
\bibinfo{author}{\bibfnamefont{M.}~\bibnamefont{Stanek}},
  \bibinfo{author}{\bibfnamefont{L.}~\bibnamefont{Wierzbicki}},
  \bibnamefont{and}
  \bibinfo{author}{\bibfnamefont{M.}~\bibnamefont{Leonowicz}},
  \bibinfo{journal}{Arch. Metall. Mater.} \textbf{\bibinfo{volume}{55}},
  \bibinfo{pages}{571} (\bibinfo{year}{2010}).

\bibitem[{\citenamefont{K\"ormann et~al.}(2010)\citenamefont{K\"ormann, Dick,
  Hickel, and Neugebauer}}]{Koermann-2010}
\bibinfo{author}{\bibfnamefont{F.}~\bibnamefont{K\"ormann}},
  \bibinfo{author}{\bibfnamefont{A.}~\bibnamefont{Dick}},
  \bibinfo{author}{\bibfnamefont{T.}~\bibnamefont{Hickel}}, \bibnamefont{and}
  \bibinfo{author}{\bibfnamefont{J.}~\bibnamefont{Neugebauer}},
  \bibinfo{journal}{Phys. Rev. B} \textbf{\bibinfo{volume}{81}},
  \bibinfo{pages}{134425} (\bibinfo{year}{2010}).

\bibitem[{\citenamefont{Yuhasz et~al.}(2009)\citenamefont{Yuhasz, Schlagel,
  Xing, Dennis, McCallum, and Lograsso}}]{Yuhasz-2009}
\bibinfo{author}{\bibfnamefont{W.~M.} \bibnamefont{Yuhasz}},
  \bibinfo{author}{\bibfnamefont{D.~L.} \bibnamefont{Schlagel}},
  \bibinfo{author}{\bibfnamefont{Q.}~\bibnamefont{Xing}},
  \bibinfo{author}{\bibfnamefont{K.~W.} \bibnamefont{Dennis}},
  \bibinfo{author}{\bibfnamefont{R.~W.} \bibnamefont{McCallum}},
  \bibnamefont{and} \bibinfo{author}{\bibfnamefont{T.~A.}
  \bibnamefont{Lograsso}}, \bibinfo{journal}{J. Appl. Phys.}
  \textbf{\bibinfo{volume}{105}}, \bibinfo{pages}{07A921}
  (\bibinfo{year}{2009}).

\bibitem[{\citenamefont{Yuhasz et~al.}(2010)\citenamefont{Yuhasz, Schlagel,
  Xing, McCallum, and Lograsso}}]{Yuhasz-2010}
\bibinfo{author}{\bibfnamefont{W.}~\bibnamefont{Yuhasz}},
  \bibinfo{author}{\bibfnamefont{D.}~\bibnamefont{Schlagel}},
  \bibinfo{author}{\bibfnamefont{Q.}~\bibnamefont{Xing}},
  \bibinfo{author}{\bibfnamefont{R.}~\bibnamefont{McCallum}}, \bibnamefont{and}
  \bibinfo{author}{\bibfnamefont{T.}~\bibnamefont{Lograsso}},
  \bibinfo{journal}{J. Alloys Compd.} \textbf{\bibinfo{volume}{492}},
  \bibinfo{pages}{681} (\bibinfo{year}{2010}).

\bibitem[{\citenamefont{Krenke et~al.}(2016)\citenamefont{Krenke, Çakır,
  Scheibel, Acet, and Farle}}]{Krenke-2016}
\bibinfo{author}{\bibfnamefont{T.}~\bibnamefont{Krenke}},
  \bibinfo{author}{\bibfnamefont{A.}~\bibnamefont{Çakır}},
  \bibinfo{author}{\bibfnamefont{F.}~\bibnamefont{Scheibel}},
  \bibinfo{author}{\bibfnamefont{M.}~\bibnamefont{Acet}}, \bibnamefont{and}
  \bibinfo{author}{\bibfnamefont{M.}~\bibnamefont{Farle}}, \bibinfo{journal}{J.
  Appl. Phys.} \textbf{\bibinfo{volume}{120}}, \bibinfo{pages}{243904}
  (\bibinfo{year}{2016}).

\bibitem[{\citenamefont{{\c{C}}ak{\i}r
  et~al.}(2016)\citenamefont{{\c{C}}ak{\i}r, Acet, and Farle}}]{cakir-2016}
\bibinfo{author}{\bibfnamefont{A.}~\bibnamefont{{\c{C}}ak{\i}r}},
  \bibinfo{author}{\bibfnamefont{M.}~\bibnamefont{Acet}}, \bibnamefont{and}
  \bibinfo{author}{\bibfnamefont{M.}~\bibnamefont{Farle}},
  \bibinfo{journal}{Sci. Rep.} \textbf{\bibinfo{volume}{6}}, \bibinfo{pages}{1}
  (\bibinfo{year}{2016}).

\bibitem[{\citenamefont{Çakır et~al.}(2017)\citenamefont{Çakır, Acet,
  Wiedwald, Krenke, and Farle}}]{Cakir-2017}
\bibinfo{author}{\bibfnamefont{A.}~\bibnamefont{Çakır}},
  \bibinfo{author}{\bibfnamefont{M.}~\bibnamefont{Acet}},
  \bibinfo{author}{\bibfnamefont{U.}~\bibnamefont{Wiedwald}},
  \bibinfo{author}{\bibfnamefont{T.}~\bibnamefont{Krenke}}, \bibnamefont{and}
  \bibinfo{author}{\bibfnamefont{M.}~\bibnamefont{Farle}},
  \bibinfo{journal}{Acta Mater.} \textbf{\bibinfo{volume}{127}},
  \bibinfo{pages}{117} (\bibinfo{year}{2017}).

\bibitem[{\citenamefont{Entel et~al.}(2018)\citenamefont{Entel, Gruner, Acet,
  {\c{C}}ak{\i}r, Arr{\'o}yave, Duong, Sahoo, F{\"a}hler, and
  Sokolovskiy}}]{Entel-2018}
\bibinfo{author}{\bibfnamefont{P.}~\bibnamefont{Entel}},
  \bibinfo{author}{\bibfnamefont{M.~E.} \bibnamefont{Gruner}},
  \bibinfo{author}{\bibfnamefont{M.}~\bibnamefont{Acet}},
  \bibinfo{author}{\bibfnamefont{A.}~\bibnamefont{{\c{C}}ak{\i}r}},
  \bibinfo{author}{\bibfnamefont{R.}~\bibnamefont{Arr{\'o}yave}},
  \bibinfo{author}{\bibfnamefont{T.}~\bibnamefont{Duong}},
  \bibinfo{author}{\bibfnamefont{S.}~\bibnamefont{Sahoo}},
  \bibinfo{author}{\bibfnamefont{S.}~\bibnamefont{F{\"a}hler}},
  \bibnamefont{and} \bibinfo{author}{\bibfnamefont{V.~V.}
  \bibnamefont{Sokolovskiy}}, \bibinfo{journal}{Energy Technol.}
  \textbf{\bibinfo{volume}{6}}, \bibinfo{pages}{1478} (\bibinfo{year}{2018}).

\bibitem[{\citenamefont{Sokolovskiy et~al.}(2019)\citenamefont{Sokolovskiy,
  Gruner, Entel, Acet, {\c{C}}ak{\i}r, Baigutlin, and
  Buchelnikov}}]{Sokolovskiy-2019}
\bibinfo{author}{\bibfnamefont{V.~V.} \bibnamefont{Sokolovskiy}},
  \bibinfo{author}{\bibfnamefont{M.~E.} \bibnamefont{Gruner}},
  \bibinfo{author}{\bibfnamefont{P.}~\bibnamefont{Entel}},
  \bibinfo{author}{\bibfnamefont{M.}~\bibnamefont{Acet}},
  \bibinfo{author}{\bibfnamefont{A.}~\bibnamefont{{\c{C}}ak{\i}r}},
  \bibinfo{author}{\bibfnamefont{D.~R.} \bibnamefont{Baigutlin}},
  \bibnamefont{and} \bibinfo{author}{\bibfnamefont{V.~D.}
  \bibnamefont{Buchelnikov}}, \bibinfo{journal}{Phys. Rev. Materials}
  \textbf{\bibinfo{volume}{3}}, \bibinfo{pages}{084413} (\bibinfo{year}{2019}).

\end{thebibliography}

\end{document}